\documentclass[a4paper, usenatbib]{mnras}

\usepackage{graphicx}
\usepackage{xcolor}
\usepackage{amsmath}
\usepackage{subfig}
\usepackage{longtable}
\usepackage{array, booktabs, ltablex, makecell, threeparttablex}
\usepackage{ragged2e}
\usepackage[labelfont=bf,font=small]{caption}
\usepackage{float}

\makeatletter
 \def\@textbottom{\vskip \z@ \@plus 1pt}
 \let\@texttop\relax
\makeatother

\def\LaTeX{L\kern-.36em\raise.3ex\hbox{a}\kern-.15em
    T\kern-.1667em\lower.7ex\hbox{E}\kern-.125emX}

\graphicspath{{./}{Figures/}}

\title[Swift/BAT AGN Variability]{Investigating Non-linear and Stochastic Hard X-ray Variability of Active Galactic Nuclei using Recurrence Analysis}

\author[R. A. Phillipson et al.]
  {R.\,A.~Phillipson\href{https://orcid.org/0000-0001-6891-7091}{\textcolor[HTML]{A6CE39}{\includegraphics[scale=0.5]{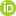}}},$^{1,2}$\thanks{E-mail: raphilli@uw.edu}
  M.\,S.~Vogeley\href{https://orcid.org/0000-0001-7416-9800}{\textcolor[HTML]{A6CE39}{\includegraphics[scale=0.5]{ORCID_icon.png}}},$^3$
  P.\,T.~Boyd\href{https://orcid.org/0000-0003-0442-4284}{\textcolor[HTML]{A6CE39}{\includegraphics[scale=0.5]{ORCID_icon.png}}}$^4$\\
  $^1$University of Washington, Department of Astronomy, 3910 15th Avenue NE, Seattle, WA 98195, USA\\
  $^2$Villanova University, Department of Physics, Villanova, PA 19085, USA\\
  $^3$Drexel University, Department of Physics, 3141 Chestnut St, Philadelphia, PA 19104, USA\\
  $^4$Astrophysics Science Division, NASA Goddard Space Flight Center, Greenbelt, MD 20771, USA}
\date{Accepted 2022 November 18. Received 2022 November 14; in original form 2022 September 17}
\pubyear{2022}

\begin{document}
\label{firstpage}
\pagerange{\pageref{firstpage}--\pageref{lastpage}}
\maketitle

\begin{abstract}

We present results of recurrence analysis of 46 active galactic nuclei (AGN) using light curves from the 157-month catalog of the Swift Burst Alert Telescope (BAT) in the 14-150 keV band. We generate recurrence plots and compute recurrence plot metrics for each object. We use the surrogate data method to compare all derived recurrence-based quantities to three sets of stochastic light curves with identical power spectrum, flux distribution, or both, in order to determine the presence of determinism, non-linearity, entropy, and non-stationarity. We compare these quantities with known physical characteristics of each system, such as black hole mass, Eddington ratio, and bolometric luminosity, radio loudness, obscuration, and spectroscopic type. We find that almost all AGN in this sample exhibit substantial higher-order modes of variability than is contained in the power spectrum, with approximately half exhibiting nonlinear or non-stationary behavior. We find that Type 2 AGN are more likely to contain deterministic variability than Type 1 AGN while the same distinction is not found between obscured and unobscured AGN. The complexity of variability among Type 1 AGN is anticorrelated with Eddington ratio, while no relationship is found among Type 2 AGN. The connections between the recurrence properties and AGN class suggest that hard X-ray emission is a probe of distinct accretion processes among classes of AGN, which supports interpretations of changing-look AGN and challenges the traditional unification model that classifies AGN only on viewing angle. 

\end{abstract}


\section{Introduction}

Active Galactic Nuclei (AGN) are highly luminous sources with variations in luminosity on timescales from hours up to years \citep{Smith1963, Matthews1963}. Due to the dramatic variability in luminosity over short timescales and the compact size of their emitting regions, it is widely accepted that AGN are powered by the accretion of matter onto a central black hole \citep{Salpeter1964, LyndenBell1969} where gravitational potential energy is converted into radiative output \citep{Fabian1979, Rees1982} with varying degrees of radiative efficiency depending on the structure of the accretion flow. The basic picture of an AGN consists of a supermassive black hole surrounded by an optically thick plasma emitting strongly in the UV/optical and soft X-ray, with an innermost, presumably hot and optically thin, coronal plasma producing the hard X-ray emission. In the ``unification'' picture of AGN classification, the two broad categories of AGN, Type 1 and Type 2, are thought to be distinct from each other primarily in viewing angle, such that the latter class of objects are largely obscured from our direct line of sight by cold absorbing material \citep{Antonucci1993}. A sub-set of AGN exhibit highly relativistic outflows of energetic particles along the rotation axis of the system, resulting in extended and collimated radio-emitting jets \citep{Ulrich1997}. The variation and emission of the accretion disk and the existence of a jet are likely intrinsically connected and there is evidence that a fundamental plane of black hole activity \citep{Merloni2003} correlates the radio and X-ray luminosities of AGN with black hole mass even among radio quiet AGN. In lower luminosity AGN, the optical variability in the accretion disk may be illuminated and thus driven by variability of the X-ray emission from the coronal region \citep{Leighly2004, Luo2015}. The evolution of the X-ray emission thus offers a particularly powerful window into the structure of the accretion flow and the relationship between different emitting regions in the system.

There is a prolific history of X-ray variability studies of AGN with a number of important results connecting physical characteristics to specific X-ray timing properties. The power spectral density (PSD) function of AGN X-ray light curves (primarily observed by the Rossi X-ray Timing Explorer in the X-ray bands less than 10 keV) can be well fit by a broken power-law in which the slope flattens towards lower frequencies. The so-called break frequency where the slope changes correlates to the black hole mass and accretion rate (e.g., \citealt{Uttley2002, McHardy2004, McHardy2006}). In the harder X-ray bands, AGN spectra show different variability characteristics between continuum and reflection components, where the latter manifests as a peak above 10 keV, called the Compton hump, and varies separately from the intrinsic and more variable power law at lower energies \citep{Sobolewska2009, Miniutti2007}. 

The \textit{Swift} Burst Alert Telescope (BAT) provides monitoring in the hard X-ray between 14 and 195 keV, providing a relatively unbiased sample of AGN that is minimally affected by the absorption that can impact lower energy bandwidths \citep{Ricci2017, Koss2017}, and covering the energy range above 10 keV in which variability components from both the reflection and emission appear. Using the 9-month Swift/BAT catalog, \cite{Beckmann2007} find that absorbed Type 2 AGN show more variability than unabsorbed Type 1 AGN and an anti-correlation between luminosity and variability using structure function analysis and excess variance. A follow-up study by \cite{Soldi2014} similarly find that a majority of the AGN in the 58-month catalog exhibit significant variability on month to year time scales and, in particular, radio loud sources are the most variable, with Type 2 AGN slightly more variable than Type 1 AGN. \citet{Shimizu2013} find for 40 AGN in the 58-month \textit{Swift}/BAT catalog that all but one source PSD could be fit by a simple power law with slope $\alpha \sim -0.8$, similar to the spectrum measured in the 2-10 keV regime for low frequencies (below the break frequency).

Recent work in the study of the optical light curves of AGN exemplifies how methods from nonlinear time series analysis can recover the same variability information as contained in the PSD and additionally probe traces of nonlinear structure (\citealt{Phillipson2020}). Applications of the same method have been used to study the stochastic and nonlinear modes of Galactic X-ray binaries (\citealt{Sukova2016, Phillipson2018}) and suggests that we test for similar behavior in the long-term monitoring of AGN.

In this study we will test for nonlinear behavior in the hard X-ray light curves of AGN provided by the Swift/BAT 157-month catalog using recurrence analysis. We will attempt to distinguish between stochastic and nonlinear classes of variability specific to other AGN physical characteristics primarily provided by the optical spectroscopic analyses performed by the \textit{Swift}/BAT AGN Spectroscopic Survey (BASS; \citealt{Koss2017}). \cite{Giustini2019} provide a model of AGN high-energy spectral states (Giustini-Proga, hereafter), similar to soft-energy/high-intensity and hard-energy/low-intensity states of Galactic X-ray binaries, based on the accretion rate, X-ray luminosity, and black hole mass of an AGN (see Sec.~\ref{sec:GPclass} for the specific criteria defining each class). With the physical characteristics of AGN and X-ray variability features in hand, the \textit{Swift}/BAT sample enables us to similarly test whether the differing Giustini-Proga spectral states correspond to specific variability characteristics as seen in Galactic black hole X-ray light curves.

This paper is organized as follows: in Sec.~\ref{sec:methods} we introduce the concept of a recurrence plot, which is the organization of all the flux differences related in time into a 2D matrix, and the surrogate data method for determining significance of our results. In Sec.~\ref{sec:data} we discuss the construction of the \textit{Swift}/BAT light curves from the 157-month catalog, our choice of binning, and selection criteria for the AGN sample in this study. In Sec.~\ref{sec:analysis} we present our results, including the generation of recurrence plots for all 46 selected AGN, the characteristic behaviors evident in the recurrence plots, and recurrence plot quantities that correlate with nonlinear, deterministic, and stochastic behavior. Finally, in Sec.~\ref{sec:results} we compare the significance of the behaviors probed by the recurrence properties to the physical characteristics of each AGN including spectroscopic type, obscuration, radio properties, Giustini-Proga class, black hole mass, Eddington ratio, and luminosity. We present our conclusions and discussion of relationships in Sec.~\ref{sec:conclusions}. We have also included three appendices available online that provide additional details to some of the methods utilized in the analysis. Appendix A walks through generating a recurrence plot of an example light curve (the bright, well-studied X-ray binary, Her X-1). Appendix B describes our assessment of the binning and signal-to-noise requirements for recurrence analysis of the \textit{Swift}/BAT light curves. Appendix C provides a discussion of the systematics we discovered in the \textit{Swift}/BAT light curves and their impact on recurrence analysis. The online supplemental figures contain all of the light curves and recurrence plots for each source in the study.

\section[Methods]{Methods: The Recurrence Plot and Surrogate Data}\label{sec:methods}

\subsection{The Recurrence Plot}

The concept of recurrent behavior in time series was first introduced by Henri Poincar\'{e}, with visualization of recurrences in the form of Poincar\'{e} plots, or return maps \citep{Hilborn}, and is one of the foundational elements of dynamical systems. Recurrence Plots (RPs) were introduced by \cite{Eckmann1987} as a more general means to visualize the recurrences of trajectories within dynamical systems. RPs provide qualitative information about the behavior of a time series and its underlying system, including indications of random, periodic, or chaotic behavior. Measures that quantify the structures present in RPs were introduced by \cite{Webber1994}, collectively called Recurrence Quantification Analysis (RQA), and subsequently applied to various fields including Mathematics, Geology, and Physiology \citep{Gao2000, Marwan2002, Zbilut2002} and others. RQA has been utilized in the study of the stability of terrestrial planets \citep{Asghari2004}, in the distinction of chaotic and stochastic behavior in the X-ray variability of microquasars \citep{Sukova2016}, and in the confirmation of a quasi-periodic oscillation and the distinction of deterministic versus stochastic behavior in two \textit{Kepler}-monitored AGN \citep{Phillipson2020}. For an extensive overview of the history of RPs, RQA, and their applications, see the seminal review by \cite{Marwan2007}.

Following the notation of \cite{Marwan2007}, suppose we have a dynamical system represented by the trajectory $\vec{x}_i$ for $i=1,...,N$ in a $d$-dimensional phase space (for example, for a classical equation of motion like a pendulum, a 3d phase space is constructed from position, velocity, and acceleration) where $N$ is the number of measured points, $\vec{x}_i$. For dynamical systems in which there is only one observable (i.e., scalar time series), a representative phase space can be constructed (e.g., via the time delay method, used in this study). The recurrence matrix is then defined as
\begin{equation}\label{eq:rp_mat1}
\mathbf{R}_{i,j}(\epsilon) = \Theta(\epsilon - ||\vec{x}_i - \vec{x}_j ||) \,\, \text{for} \,\, i,j = 1, ... , N,
\end{equation}
where $\epsilon$ is a threshold distance, and $\Theta(\cdot)$ is the Heaviside function. For states that persist in an $\epsilon$-neighborhood, i.e. return to within a threshold distance of a previous state, the following condition holds:
\begin{equation}\label{eq:condition1}
\vec{x}_i \approx \vec{x}_j \Leftrightarrow \mathbf{R}_{i,j} = 1.
\end{equation}
RPs are the graphical representation of the binary recurrence matrix, Eq.~\ref{eq:rp_mat1}, where a color represents each entry of the matrix (e.g. a black dot for unity and empty for zero). By convention, the axes increase in time rightwards and upwards. The RP is also symmetric about the main diagonal, called the line of identity (LOI). The patterns observed in the RP effectively map out phase space trajectories that are recurrent and are thus representative of the underlying dynamical system. The quantification of these structures, particularly the diagonal lines, can be used to distinguish classes of dynamical systems. 

There are two considerations for generating a RP: the construction of phase space for univariate data and the determination of the threshold, $\epsilon$. The phase space of the time series is commonly reconstructed via a time delay embedding \citep{Takens1981}, which is a method that reconstructs the phase space from a single observable by sampling from the 1D time series at evenly spaced intervals. We use the time delay method in this study (discussed in Sec.~\ref{sec:data}; a detailed discussion of the method can be found in \citealt{Phillipson2020}). As a rule of thumb, $\epsilon$ should not exceed approximately 10 percent the maximum phase space diameter \citep{Zbilut1992}. More importantly, dynamical invariants appear over a range of thresholds and therefore it is best practice to vary the threshold ranging from a value larger than the observational noise up to saturation of the RP. 

The structures (or lack thereof) present in RPs correlate to specific dynamical behavior of the system and can thus be used to classify time series based on their RPs. For example, white noise will produce randomly distributed black dots, periodic signals will produce diagonal lines offset from the LOI by the period, non-stationarity gives rise to progressively faded regions at the corners of the RPs, and laminar states manifest as horizontal or vertical lines \citep{Marwan2007}. The distribution of diagonal lines can be related to the correlation integral, correlation dimension, and topological entropy, which is a measure of the complexity and predictability of the system, as well as measures for determinism versus randomness \citep{Faure1998}. Appendix A, available online, walks through an example of generating a RP with \textit{Swift}/BAT observations of the X-ray Binary, Her X-1.

\subsection{The Surrogate Data Method}

A method for determining the significance of recurrent phenomena present in a RP and for distinguishing between stochastic, deterministic, and nonlinear behavior is the Surrogate Data Method (\citealt{Theiler1992}). The surrogate data method is a constrained realizations approach (\citealt{Theiler1996}) in which a null hypothesis is formed, and data are tested against this hypothesis by comparing to a set of surrogate time series generated directly from the data itself (rather than from an assumed statistical model). Using a specific test statistic, if the data performs significantly differently from the set of surrogates, then we can rule out the null hypothesis as an explanation for the observables of interest. 

In this study we will generate three types of surrogate data: ``shuffled'' corresponding to uncorrelated noise preserving the distribution of the source, ``phase'' corresponding to correlated noise preserving the power spectrum of the data, and ``IAAFT'' corresponding to iterative amplitude adjusted fourier transform surrogates preserving both the power spectrum and distribution. An example of generating these types of surrogates is discussed in Appendix A, available online.

\subsection{Test Statistics from the RP and Light Curves}

In this study, we will use the method of surrogate data and three measures calculated from the RP to detect traces of nonlinearity, determinism, and high information entropy as well as two measures from traditional surrogate data testing. If a null hypothesis is rejected via a discriminating statistic, we can only say that the process consistent with the null hypothesis is not a valid model for the system. If the discriminating statistic for the data is indistinguishable from surrogates, we cannot say that the null hypothesis is true, only that we failed to reject it. Thus, it is important to use multiple discriminating statistics to probe different features of the data \citep{Kugiumtzis2001}.

\subsubsection{Nonlinearity ($L_{max}$) Test Statistic}
The first measure is the longest diagonal line length ($L_{max}$) (as used in the example in Appendix A, available online), for the detection of nonlinearity. $L_{max}$ is a proxy for the $K_2$ entropy, which is related to the maximum Lyapunov exponent that describes the rate of decay of nearby initial conditions for the geometric object (attractor) manifested from a set of differential equations. This measure has been used as a first-pass indication of nonlinearity, approximating results from calculating the $K_2$ entropy outright in AGN monitored by \textit{Kepler} \citep{Phillipson2020}, when compared against the IAAFT surrogates, and as a probe of nonlinearity in XRBs \citep{Sukova2016}. When significant against the phase or IAAFT surrogates, $L_{max}$ gives an indication that higher-order moments beyond the second-order metrics contained in a PSD are responsible for the observed nonlinearity.

\subsubsection{Determinism (DET) Test Statistic}
The second measure computed from the RP is the ``determinism'', or DET. The DET measure is defined as the fraction of recurrence points that are part of diagonal lines versus total number of recurrence points. DET gives an indication of regularity, typically indications of periodicities; these are periodicities reflected in phase space, and thus are likely tracers of unstable periodic orbits (UPOs; \citealt{Gilmore1998}). UPOs exist in periodic, linear, and nonlinear systems and their organization can be useful as probes of a specific class of differential equations describing the physical process (e.g., the damped and driven Duffing oscillator UPOs are identical to those of a neutron star X-ray binary in \citealt{Phillipson2018}). In this context, DET probes the presence of possible UPOs, not necessarily measures of nonlinearity, and is thus more associated with the determinism of a system (and whether it rises above the stochastic signal). Significance against the phase or IAAFT surrogates means these UPOs are best described by higher-order moments than the PSD.

\subsubsection{Entropy ($L_{entr}$) Test Statistic}
The third measure calculated from the RP is the Shannon entropy ($L_{entr}$; \citealt{Shannon1948, Shannon1951}), which is a measure used most often in information theory. It measures the information --- defined as the uncertainty --- associated with the physical process described by the probability distribution (also known as the information entropy; higher values correspond to more unpredictability and uncertainty in the time series distribution). Minimum uncertainty, and thus perfect order, corresponds to a minimum in the Shannon entropy while maximum uncertainty, or total randomness, is associated with high Shannon entropy. Put another way, the entropy is the average number of bits needed to optimally encode independent draws of a variable following a probability distribution. When significant against the surrogates (e.g., systematically lower than the ensemble of surrogates), then we can say that there is a physical process distinct from the null hypothesis describing the surrogate type \citep{Thuraisingham2019}. Systematically higher quantities correspond to light curves with high entropy and uncertainty in the physical process, which may also indicate chaos. Whether systematically higher or lower than the surrogates, such a distinction would indicate that a Gaussian process is not an appropriate assumption for the light curve.

\subsubsection{The Surrogate Data Method Test Statistics}

Finally, we will compare against two traditional test statistics in the surrogate data method \citep{Schreiber2000} that do not utilize a RP. The first is the time reversal asymmetry statistic. A time series is said to be reversible if its statistical properties are invariant with respect to the direction of time. Non-equilibrium systems and dynamics resulting from non-conservative forces lead to time irreversibility \citep{Lamb1998}. In contrast, all Gaussian processes (and all static transformations of a linear Gaussian process) are time-reversible since their joint distributions are determined by the symmetric covariance function \citep{Weiss1975}. A system that is non-linear will lead to time irreversibility, or poor forecasting of the time-reversed data. If reversibility can be rejected, then all static transformations of linear stochastic processes can be excluded as a model for the original time series \citep{Diks}. The PSD does not contain information about the direction of time and is therefore insufficient for distinguishing linear stochastic and other related processes. 

We compute the time reversibility asymmetry statistic for the data and each of the surrogates using the \textit{TISEAN} package. If the data is significantly different from the surrogates, then the null hypothesis of any static transformation of a linear stochastic process generating the time series can be rejected. Though time reversal asymmetry detects non-linearity, it does not provide information about the nature of the non-linearity, i.e. whether it originates from a deterministic or stochastic mechanism. For example, nonlinear shot noise processes are inherently stochastic but exhibit canonical nonlinearity such as time irreversibility \citep{Weiss1975, Eliazar2005}.

The second test statistic is the simple non-linear prediction algorithm \citep{Kantz}. This method proposes that, if we have observed a system for some time, there will be states in the past which are arbitrarily close to the current state. A ``state'' is considered to be some segment of the time series embedded in phase space. We can therefore use a localised segment of the embedded time series to predict, or forecast, future behaviour. 

The \textit{TISEAN} package provides a zeroth-order prediction of the dynamics where we use a locally constant predictor to forecast the future of the time series. If the resulting $rms$-error in the prediction algorithm \citep{Farmer} is smallest for the original time series than in its surrogates, the null hypothesis is rejected \citep{Schreiber2000}. If the null hypothesis is not rejected, this implies that predictability is not significantly reduced by destroying possible non-linear structure.

\section{Data Reduction and Sample Selection}\label{sec:data}

\subsection{Light Curve Preparation}

The Neil Gehrels Swift Burst Alert Telescope \citep{Gehrels2004, Barthelmy2005} has been observing the hard X-ray sky (14-195 keV), primarily searching for gamma-ray bursts, since its operation start in November 2004. \textit{Swift}/BAT uses a 5200 cm$^2$ coded-aperture mask \citep{Dicke1968} above an array of detectors to produce a wide field of view of the sky up to $\sim$1 str. While not observing gamma-ray bursts, the BAT is continuously observing the sky in five minute snapshot images that are converted into light curves for X-ray sources in the field of view \citep{Tueller2010}. Multiple catalog releases of the resulting multi-year light curves have been periodically released by the \textit{Swift} team \citep{Markwardt2005, Tueller2008, Tueller2010, Baumgartner2013, Oh2018}. For this study, we use the 157-month catalog (Lien et al., in preparation), for which derived properties including black hole mass, accretion rate, and bolometric luminosity are also provided in the \textit{Swift}/BAT spectroscopic sample (DR1; \citealt{Koss2017}) for 74 percent of the 836 detected AGN in the 70-month catalog.  

The light curves provided by the BAT team contain individual background-subtracted count rates in eight different energy bands: 14-20, 20-24, 24-35, 35-50, 50-75, 75-100, 100-150, and 150-195 keV, plus a total count rate for the entire 14-195 keV band. Each datapoint consists of a single 5 minute snapshot image after cleaning by the BAT pipeline (see \citealt{Tueller2010} for details) that includes anomaly filtering, corrections due to bright sources and pattern noise, and background subtraction for each detector that contribute to the individual energy bands. For our analysis, we obtain the snapshot light curves from the HEASARC\footnote{https://swift.gsfc.nasa.gov/results/bs157mon/}. A tangential motivation of this study is to compare our results to PSD analyses, such as that performed by \citet{Shimizu2013}, to determine how recurrence properties illuminate higher-order structure in the light curves. Following \citet{Shimizu2013}, who found that the 150-195 keV band low signal-to-noise did not increase variability information, we add together the first seven bands (14-150 keV) to create the total band raw light curve. \citet{Shimizu2013} re-binned the raw light curves into 5 day intervals to counteract the white noise level that overwhelms at high temporal frequencies due to Poisson noise. We verify this binning requirement for our analysis in Appendix B, available online. The raw light curves are binned into time bins with 5 day widths using a weighted average (count rates from each snapshot falling within the time bin weighted by their respective errors). Any time bin that either contains a fractional exposure time of less than 2 per cent of the 5 day time bin or gaps due to the BAT pipeline temporal filtering process were filled via a univariate spline linear interpolation connecting adjacent bins. We found that other interpolation methods, such as filling gaps with a random number drawn from the distribution of the light curve, exacerbated systematics evident in the light curve (further details on these systematics and the binning requirement are found in Appendix B, available online). For all light curves, interpolation did not exceed 7 per cent of the light curve.

We are interested in comparing the results of recurrence analysis to previous studies of the \textit{Swift}/BAT AGN for which we can test specific hypotheses with other well-established variability techniques such as the PSD and structure function. In particular, we seek to determine whether there are differing recurrence properties between Type 1 and Type 2 AGN, and between obscured and unobscured AGN, and whether there is a dependence of certain recurrence properties with luminosity. We thus consider the combined samples studied by \cite{Beckmann2007} and \cite{Shimizu2013}. We note that there have been extensive studies of the X-ray PSDs of AGN, and their counterparts, black hold X-ray binaries, in which the long timescales evident in AGN X-ray PSDs are best fitted by a power-law of slope -1 (indeed, this is found in the sample from \citealt{Shimizu2013}). However, on timescales shorter than a `break' timescale the power-law slope steepens to greater than 2. The break timescales have been found to correlate with black hole mass and accretion rate (approximated by the Eddington ratio), as detailed in \cite{McHardy2006}. Based on the measurements presented in Table~\ref{tab:table1}, the break timescales in our sample could be predicted to be in the range of about a week. Thus, the 5-day binning is likely insufficient to resolve the variability features at high frequencies. The variability characteristics that will be uncovered with recurrence analysis will instead correspond to long-term trends in the $\sim$15 year light curves, potentially related to the global instabilities of the accretion flow process.

There may be objects contaminated by large excess variance from systematics common in the \textit{Swift}/BAT light curves over periods longer than a day (e.g. bright sources nearby that were not completely removed from the \textit{Swift}/BAT team pipeline and background that manifests as pattern noise) relative to intrinsic variability. \cite{Liu2020} found that the excess variance --- defined as $\sigma^2_{xs} = S^2 - \overline{\sigma^2_{err}}$, where $S^2$ is the variance of the light curve, and $\sigma_{err}$ is the measurement error --- of AGN closely mimic that of constant hard X-ray sources such as galaxy clusters for $\sigma^2_{xs} < 1.5 \times 10^{-7}$ (where measurement errors and flux are measured in counts s$^{-1}$). We therefore exclude those AGN that \cite{Liu2020} found did not meet this threshold. 7 AGN from the combined sample did not meet this criterion (3C 120, 3C 390.3, 4C +71.07, Mrk 3, NGC 1142, NGC 3227, and NGC 4051). Finally, we include PKS 0018-19, an unobscured Type 1.9 AGN, in order to increase the sub-sample of unobscured AGN in our selection.

The resulting sample consists of 46 Seyfert 1s, Seyfert 2s, and blazars, listed in Table~\ref{tab:table1}. The spectroscopic type, redshift, and obscuration properties for each object are determined by \cite{Koss2017} (BASS DR1) and the signal-to-noise directly from the HEASARC catalog. The estimated black hole mass, bolometric luminosity, and Eddington ratio for many of the sources were additionally estimated by \cite{Koss2017}; for those measurements not contained in the BASS DR1, we include values from the literature (references in table). 

For the majority of sources, the \cite{Kellermann1989} quantity $R0 = S_{\nu,5Ghz}/S_{B}$ is used to identify the radio properties of the AGN, where $R0\sim10$ is typically considered the boundary between radio-loud and radio-quiet objects. These values are found in the literature for most sources in our sample (e.g., \citealt{Smith2020}, \citealt{Fischer2021}, \citealt{Panessa2022}, \citealt{Laor2008}, and \citealt{Rush1996}), as identified in Table~\ref{tab:table1}. 9 sources are identified as radio-quiet by \cite{Xu1999}, which determine distinct radio classes in the correlations between the 5 GHz radio luminosity relative to the [OIII+] line luminosity. For the three sources that do not explicitly appear in the literature (4U 0557-38, ESO 297-018, and ESO 506-027), we use the \cite{Gupta2018} definition, $R = F(1.4)/F(\nu,W3)$, where $F(1.4)$ and $F(\nu,W3)$ are the monochromatic fluxes at 1.4 GHz and $2.5\times10^{13}$ Hz, respectively, using historical data published in the NASA/IPAC Extragalactic Database (NED\footnote{http://ned.ipac.caltech.edu/}). Two sources did not have any radio detections (4U 1344-60 and ESO 103-035), which we also classify as radio quiet. All objects considered luminous in the radio by any of the aforementioned definitions and that also contain evidence of linear jet morphology are classified as radio loud, including the three blazars identified by \cite{Ricci2017}, which cross-matched the Swift/BAT catalog with the Roma-BZCAT 26 catalog of blazars (Massaro et al. 2015). Finally, there are certain radio quiet sources for which \cite{Smith2020} conducted radio imaging (22 GHz 1’’ JVLA) and classified the radio cores into three groups: compact, core-dominated, extended, and jet-like. Sources classified as jet-like by \cite{Smith2020} are identified in Table~\ref{tab:table1} and are subsequently grouped with the radio-loud sources in the comparative analysis performed in Sec.~\ref{sec:results}. 

The light curves of 9 AGN are displayed in Fig.~\ref{fig:AGN_LCs_select}, with the remainder in the supplemental figures available online.

Finally, we determine that a systematic signal appears with a timescale of 365 days that is sensitive to the choice of interpolation method over gaps in the light curve and additionally appears in some of the blank sky light curves generated by the \textit{Swift}/BAT team. An overview of this systematic and its discovery is detailed in Appendix C, available online.

\begin{table*} 
\centering
\begin{tabular}{ p{2.3cm}p{1cm}p{0.8cm}p{0.8cm}p{1.4cm}p{1.0cm}p{1.2cm}p{1cm}p{1cm}p{1cm} }
	\hline
		\multicolumn{10}{c}{Physical Properties of the \textit{Swift}/BAT AGN} \\
		\hline
		Object & Type   & S/N    & z      & Obscuration         & Radio & log $M_{bh}$ & log $L_x$ & log $L_{bol}$ & $\lambda_{Edd}$ \\
		& & & & & & ($M_{sol}$) & (erg/s) & (erg/s) & \\
		\hline
3C 273          & blazar & 156.8  & 0.158     & unobs.       & RL               & $^p$8.84$^{+0.11}_{-0.08}$  & 46.48 & 46.74 & 0.19\\
3C 454.3        & blazar & 40.43  & 0.859     & unobs.       & RL               & $^q$9.2$^{+0.3}_{-0.3}$ & 47.68 & 48.58 & 19.9\\
Mrk 421         & blazar & 109.5  & 0.03      & unobs.       & RL               & $^r$8.6$^{+0.2}_{-0.2}$ & 44.52 & 45.42 & 0.06\\
\hline
3C 111          & Sy 1.2 & 40.74  & 0.049     & unobs.       & RL$^b$           & 8.3$^{+0.3}_{-0.3}$     & 44.83 & 45.36 & 0.12\\
3C 382          & Sy 1.2 & 38.07  & 0.058     & unobs.       & RL$^c$           & 8.2$^{+0.3}_{-0.3}$  & 44.84 & 44.93 & 0.05\\
4C 50.55        & Sy 1.2 & 91.88  & 0.015     & obscured     & RL$^c$           & 7.6$^{+0.3}_{-0.3}$      & 44.01 & 45.68 & 0.9\\
NGC 1275        & Sy 1.5 & 50.92  & 0.019     & unobs.       & RL$^c$           & 6.5$^{+0.3}_{-0.3}$   & 43.79 & 44.14 & 0.4\\
QSO B0241+622   & Sy 1.2 & 40.69  & 0.049     & unobs.       & RL$^d$           & 8.1$^{+0.3}_{-0.3}$       & 44.72 & 44.15 & 0.009  \\
MCG+08-11-011   & Sy 1.5 & 54.83  & 0.020     & unobs.       & RQ$^{\lambda}$ $\dagger$ & 7.6$^{+0.3}_{-0.3}$       & 44.09 & 44.39 & 0.05\\
NGC 3516        & Sy 1.2 & 64.64  & 0.009     & unobs.       & RQ$^{\lambda}$ $\dagger$ & 7.39$^{+0.06}_{-0.04}$ & 43.59 & 44.07 & 0.04   \\
NGC 5548        & Sy 1.5 & 37.48  & 0.017     & unobs.       & RQ$^{\lambda}$ $\dagger$ & $^p$7.72$^{+0.02}_{-0.02}$ & 43.71 & 44.07 & 0.018  \\
4U 0557-38      & Sy 1.2 & 18.54  & 0.034     & unobs.       & RQ$^{\delta}$    & 6.9$^{+0.3}_{-0.3}$      & 43.90 & 44.25 & 0.2\\
GRS 1734-292    & Sy 1   & 33.51  & 0.021$^a$ & obscured$^a$ & RQ$^e$           & $^a$8.5$^{+0.1}_{-0.1}$  & 44.08 & 45.16 & 0.02\\
IC 4329A        & Sy 1.5 & 101.1  & 0.016     & unobs.       & RQ$^{\beta}$           & 7.84$^{+0.3}_{-0.3}$       & 44.21 & 44.19 & 0.018\\
MR 2251-178     & Sy 1.2 & 43.95  & 0.064     & unobs.       & RQ$^{\beta}$           & 8.4$^{+0.3}_{-0.3}$  & 45.01 & 45.44 & 0.04\\
Mrk 6           & Sy 1.5 & 29.72  & 0.019     & unobs.       & RQ$^g$           & $^p$8.10$^{+0.04}_{-0.04}$  & 43.69 & 44.8  & 0.04\\
Mrk 110         & Sy 1.5 & 32.07  & 0.036     & unobs.       & RQ$^h$          & $^p$7.3$^{+0.1}_{-0.1}$ & 44.22 & 44.0  & 0.04\\
Mrk 926         & Sy 1.5 & 48.79  & 0.047     & unobs.       & RQ$^{\lambda}$           & 8.6$^{+0.3}_{-0.3}$   & 44.77 & 44.81 & 0.014\\
NGC 3783        & Sy 1.2 & 68.7   & 0.009     & unobs.       & RQ$^{\beta}$           & 7.37$^{+0.08}_{-0.08}$ & 43.69 & 44.16 & 0.05    \\
NGC 4151        & Sy 1.5 & 275    & 0.003     & obscured     & RQ$^{\beta f}$   & $^p$7.56$^{+0.05}_{-0.05}$ & 43.01 & 43.83 & 0.015\\
NGC 4593        & Sy 1.0 & 35.33  & 0.008     & unobs.       & RQ$^{\beta}$           & $^p$6.88$^{+0.1}_{-0.08}$ & 43.06 & 43.89 & 0.08 \\
NGC 6814        & Sy 1.5 & 24.79  & 0.006     & unobs.       & RQ$^i$     & $^p$7.04$^{+0.06}_{-0.06}$ & 42.68 & 43.49 & 0.02   \\
\hline
Cen A           & Sy 2   & 428.7  & 0.001     & obscured     & RL$^j$           & 7.8$^{+0.2}_{-0.2}$ & 42.36 & 43.26 & 0.002\\
Cyg A           & Sy 2   & 57.79  & 0.056     & obscured     & RL$^k$           & $^s$9.40$^{+0.12}_{-0.12}$  & 45.01 & 45.91 & 0.03\\
Mrk 348         & Sy 1.9 & 73.93  & 0.015     & obscured     & RL$^l$           & 7.6$^{+0.6}_{-0.6}$       & 43.90 & 44.80 & 0.12\\
PKS 0018-19     & Sy 1.9 & 8.54   & 0.096     & unobs.       & RL$^m$           & 9.2$^{+0.3}_{-0.3}$       & 44.6  & 45.5  & 0.017   \\
NGC 2110        & Sy 2   & 98.1   & 0.007     & obscured     & RQ$^{\lambda}$ $\dagger$ & 9.3$^{+0.3}_{-0.3}$   & 43.69 & 44.59 & 0.002 \\
NGC 5728        & Sy 1.9 & 24.34  & 0.010     & obscured     & RQ$^{\lambda}$ $\dagger$ & 8.1$^{+0.3}_{-0.3}$       & 43.00 & 43.90 & 0.005  \\
4U 1344-60      & Sy 1.9 & 42.29  & 0.013     & obscured     & RQ$^{\delta}$    & $^t$7.4$^{+0.1}_{-0.1}$ & 43.60 & 44.50 & 0.09\\
Circinus Galaxy & Sy 2   & 110.71 & 0.0014    & obscured     & RQ$^{\beta}$           & $^u$6.23$^{+0.08}_{-0.08}$ & 42.09 & 42.73 & 0.03 \\
ESO 103-035     & Sy 1.9 & 46.66  & 0.013     & obscured     & RQ$^{\delta}$    & $^t$7.3$^{+0.1}_{-0.1}$  & 43.64 & 44.6  & 0.14\\
ESO 297-018     & Sy 2   & 32.94  & 0.025     & obscured     & RQ$^{\delta}$    & 8.45$^{+0.3}_{-0.3}$       & 44.00 & 44.90 & 0.02\\
ESO 506-027     & Sy 2   & 30.68  & 0.024     & obscured     & RQ$^{\delta}$    & 9.0$^{+0.3}_{-0.3}$       & 44.10 & 45.00 & 0.008\\
MCG-05-23-016   & Sy 1.9 & 97.74  & 0.008     & obscured     & RQ$^{\lambda}$           & $^v$6.3$^{+0.7}_{-0.7}$ & 43.49 & 44.30 & 0.99\\
NGC 1365        & Sy 2   & 34.64  & 0.005     & obscured     & RQ$^{\delta}$    & $^t$7.5$^{+0.1}_{-0.1}$  & 42.40 & 43.31 & 0.006\\
NGC 2992        & Sy 1.9 & 10.75  & 0.008     & unobs.       & RQ$^{\lambda}$         & $^t$7.3$^{+0.1}_{-0.1}$  & 42.51 & 43.5  & 0.01    \\
NGC 3081        & Sy 2   & 30.41  & 0.008     & obscured     & RQ$^{\lambda}$           & 8.4$^{+0.5}_{-0.5}$    & 42.84 & 43.75 & 0.002    \\
NGC 3281        & Sy 2   & 37.42  & 0.011     & obscured     & RQ$^{\beta}$           & $^t$7.2$^{+0.1}_{-0.1}$  & 43.38 & 44.28 & 0.1   \\
NGC 4388        & Sy 2   & 110.7  & 0.008     & obscured     & RQ$^{\lambda}$          & 7.0$^{+0.4}_{-0.4}$       & 43.17 & 44.07 & 0.1    \\
NGC 4507        & Sy 1.9 & 71.13  & 0.012     & obscured     & RQ$^{\beta}$           & 7.9$^{+0.4}_{-0.4}$       & 43.76 & 44.66 & 0.05    \\
NGC 4945        & Sy 2   & 79.31  & 0.0019    & obscured     & RQ$^n$           & $^w$6.15$^{+0.16}_{-0.16}$ & 42.35 & 43.04 & 0.06    \\
NGC 5252        & Sy 2   & 46.62  & 0.023     & obscured     & RQ$^{\beta}$           & 8.9$^{+0.3}_{-0.3}$       & 44.14 & 45.04 & 0.012    \\
NGC 5506        & Sy 1.9 & 99.51  & 0.006     & obscured     & RQ$^{\lambda}$           & $^t$7.7$^{+0.1}_{-0.1}$ & 43.21 & 44.12 & 0.02   \\
NGC 7172        & Sy 2   & 63.44  & 0.008     & obscured     & RQ$^o$           & 8.5$^{+0.3}_{-0.3}$   & 43.37 & 44.27 & 0.005    \\
NGC 7582        & Sy 2   & 35.02  & 0.005     & obscured     & RQ$^{\beta}$           & $^x$7.74$^{+0.08}_{-0.29}$ & 42.63 & 43.53 & 0.005    \\
XSS J05054-2348 & Sy 2   & 25.64  & 0.036     & obscured     & RQ$^{\lambda}$           & $^t$7.7$^{+0.1}_{-0.1}$ & 44.27 & 45.17 & 0.3\\
\hline
\end{tabular}
\caption[Physical Properties of the Swift/BAT AGN]{Physical Properties of the Swift/BAT AGN sorted by spectroscopic type and then my radio properties. Columns: source name, spectroscopic type, signal-to-noise (S/N; 14-195 keV), redshift (z), obscuration (where obscured is $n_H>10^{22}cm^{-2}$), radio classification (radio loud - RL, or radio quiet - RQ), black hole mass, X-ray luminosity (14-195 keV), bolometric luminosity, and Eddington ratio ($L_{bol}/L_{Edd}$), which uses the standard relation $L_{edd} = 1.26\times10^{38}\,M_{bh}$. All quantities listed are from \cite{Koss2017}. $L_{bol}$ is derived from the relationship $log(L_{bol})=1.1157$ log$(L_X)-4.228$ in cases where $L_{bol}$ is not determined by \cite{Koss2017}. Note that the reported values from Koss are not corrected for beaming in the case of the blazars. References for all other properties not obtained from \cite{Koss2017} are annotated.\\
$\dagger$Radio quiet sources with a `jet-like' core as determined by \cite{Smith2020}\\
$^{\lambda}$\citealt{Smith2020}; $^{\beta}$\citealt{Xu1999}; $^{\delta}$NASA/IPAC Extragalactic Database \\
$^a$\citealt{Tortosa2018}; $^b$\citealt{Linfield1984}, \citealt{Molina2008}; $^c$\citealt{Bassani2016}; $^d$\citealt{Lister2015};
$^e$\citealt{Laor2008}, \citealt{Marti1998}; $^f$\citealt{Williams2020}; $^g$\citealt{Kharb2014}; $^h$\citealt{Panessa2022};  
$^i$\citealt{Fischer2021}; $^j$\citealt{Bridle1984}; $^k$\citealt{Carilli1991}; $^l$\citealt{Falcke2000}; $^m$\citealt{Gupta2018};
$^n$\citealt{Healey2007}; $^o$\citealt{Thean2000}; $^p$\citealt{Bentz2015}; $^q$\citealt{Woo2002}; $^r$\citealt{Wagner2008};  $^s$\citealt{Tadhunter2003}; $^t$\citealt{Vasudevan2010}; $^u$\citealt{Greenhill2003}; $^v$\citealt{Wang2007}; $^w$\citealt{Greenhill1997};  $^x$\citealt{Wold2006} }
\label{tab:table1}
\end{table*}

\begin{figure*}
\centering
\subfloat{
	\includegraphics[width=0.9\textwidth]{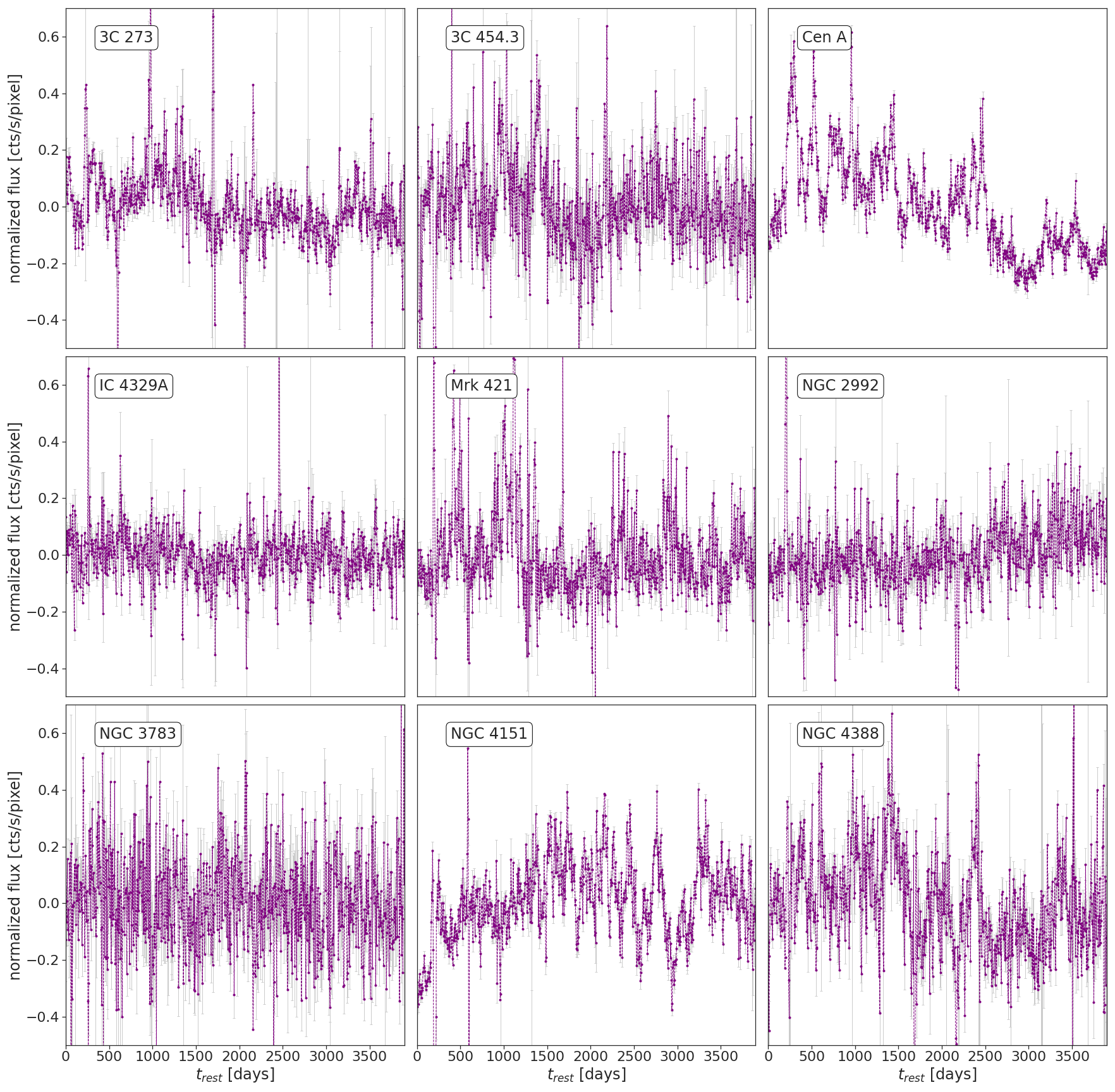}}
\caption{The 157-month 14-150 keV Swift/BAT light curves of 9 AGN, binned to 5 days. The flux is mean-subtracted and normalized for plotting purposes. The 9 AGN: 3C 273, 3C 454.3, Cen A, IC 4329A, Mrk 421, NGC 2992, NGC 3783, NGC 4151, and NGC 4388. The remaining light curves from the sample in Table~\ref{tab:table1} are contained in the supplemental figures available online.}
\label{fig:AGN_LCs_select} 
\end{figure*}

\subsection{Recurrence Plot Preparation}

In order to generate a RP for each source in Table 1, we first must determine the proper embedding parameters to cast each light curve into phase space for the computation of Eq.~\ref{eq:rp_mat1}. For the determination of the appropriate time delay to ensure minimal effects from linear or nonlinear correlations in the embedding space, we use the first minimum in the mutual information (\citealt{Fraser1986, Pompe1993}) if a pronounced minimum is evident or the autocorrelation time otherwise. The time delays range between 10 days and 85 days for this sample. 
For the selection of an appropriate embedding dimension, we use the averaged false nearest neighbors (AFNN) method \citep{Cao1997}. When an embedding dimension is too low, points that are not close to each other on the underlying attractor are projected close together and become ‘false neighbors’ in the embedded phase space. The AFNN algorithm systematically increases the embedding dimension until the rate of false neighbors is no longer changing; in this case, we choose the embedding dimension once the rate of false neighbors drops below 10 per cent. For the AGN in our sample, the optimal embedding dimensions range between 4 and 6. A table of the embedding delays and dimensions for all AGN is provided in Appendix A, available online.

For the determination of significance in our analysis and for the detection of nonlinear or deterministic behavior, we use the surrogate data method. For each light curve, we generate 100 surrogates each of the shuffled, phase, and IAAFT surrogate data types. For comparison of RPs and resulting quantitative measures of the RPs (e.g. $L_{max}$), we use the same embedding parameters on the surrogates as was determined for each source light curve.

Finally, the \textit{Swift}/BAT light curves are known to contain systematics not completely removed by the cleaning algorithm performed on the snapshot images \citep{Tueller2010}. These systematics can manifest as pattern noise in the resulting light curves particularly over long time periods. We consider the 106 Blank Sky light curves provided by the \textit{Swift}/BAT transient monitor catalog\footnote{https://swift.gsfc.nasa.gov/results/transients/BAT$\_$blank.html}, containing the light curves of ``blank'' points in the sky, randomly distributed across the sky and chosen to be at least 10 arcminutes from any reported X-ray source. We treat the ensemble of 106 blank sky light curves as an additional set of surrogate time series. For each source, we apply the same embedding parameters to the set of blank sky light curves and generate RPs. 

Using the ensemble of blank sky light curves, we confirm that spurious correlations arise on timescales less than the embedding window, defined as the time delay multiplied by the embedding dimension, also known as the ``Theiler window'' \citep{Theiler1986}. All statistics derived from the RP exclude the Theiler window (the corridor of width equal to the Theiler window centered on the main diagonal of the RP) and thus these spurious features do not impact the analysis. Secondly, we find a timescale equal to one year, or 365 days, persistent in the blank sky light curves which appears irrespective of chosen embedding parameters (see Appendix C, available online, for a discussion). To be clear, these systematics do not affect the results of the analysis as the surrogate data method ensures that the surrogates carry the same systematics as the light curves and the chosen test statistics for distinguishing between determinism, nonlinearity, and entropy do not depend on the presence of a periodic systematic imprinted on the source light curves and their surrogates.

\section{Recurrence Properties of Swift/BAT AGN}\label{sec:analysis}

\subsection{General RP Features}

We generate RPs for a range of thresholds corresponding to a recurrence point density between 1 percent and 99 percent. We use a varying threshold for two reasons: (1) a system with dynamics (not just noise) will result in recurrence plot measures that are invariant to threshold over a wide range of thresholds, and thus invariance of recurrent features can be used to identify non-stochastic sources; and (2) to determine the prevalence of recurrence features for multiple thresholded RPs for each source to ensure that we did not by happenstance select a threshold with significant behavior in the source. We show the RPs for a recurrence rate of 10 percent for 9 representative sources in our sample in Fig.~\ref{fig:AGN_RPs_select}. The remaining 37 RPs of the other sources in our sample are contained in the supplemental figures available online. The richness of information present in the phase space structure of the \textit{Swift}/BAT AGN light curves is immediately evident, and distinguishable between different sources. For example, the RPs contain many short diagonal line features or small-scale diagonally oriented structures. Diagonal lines are generally indicative of determinism, especially when the diagonal lines are long and consistently spaced. For example, a sinusoid would appear as a series of diagonal lines separated by the period. Stochastic and chaotic systems both exhibit chopped up diagonal lines, which is what we see among these AGN. 

We find that several of the RPs show evidence for non-stationarity in which large swaths of white in the RP appear far from the main diagonal. Non-stationarity in this context is defined as deviant sections of the phase-space projection of the light curve, indicating times that the source is undergoing a state change or the underlying dynamics have deviated strongly from the rest of the observations. This is most evident in Cen A (Fig. ~\ref{fig:AGN_RPs_select}) and NGC 5252 (in the supplemental figures available online). 

Some of the RPs also demonstrate changes in the texture, or small-scale patterns, over time. This is most evident in NGC 4151 in which we see dense, block-like patterns in the RP at early times that slowly evolve into larger and more sparse diagonally-oriented structures. The diagonal structures in this source also appear to not be parallel to the main diagonal of the RP, which indicates a slow change in the underlying dynamics. 

Finally, many of the RPs exhibit plaid-like patterns throughout, which indicates regular deviation from the main dynamics of the light curve (e.g., lighter or white vertical and horizontal columns). As the vertical line structures can often reflect the cadence or the time delay used in the light curve and the white gaps can reflect gaps in the light curve itself, we avoid measures that depend on either the vertical or horizontal line features. 

\begin{figure*}
\centering
\subfloat{
	\includegraphics[width=0.99\textwidth]{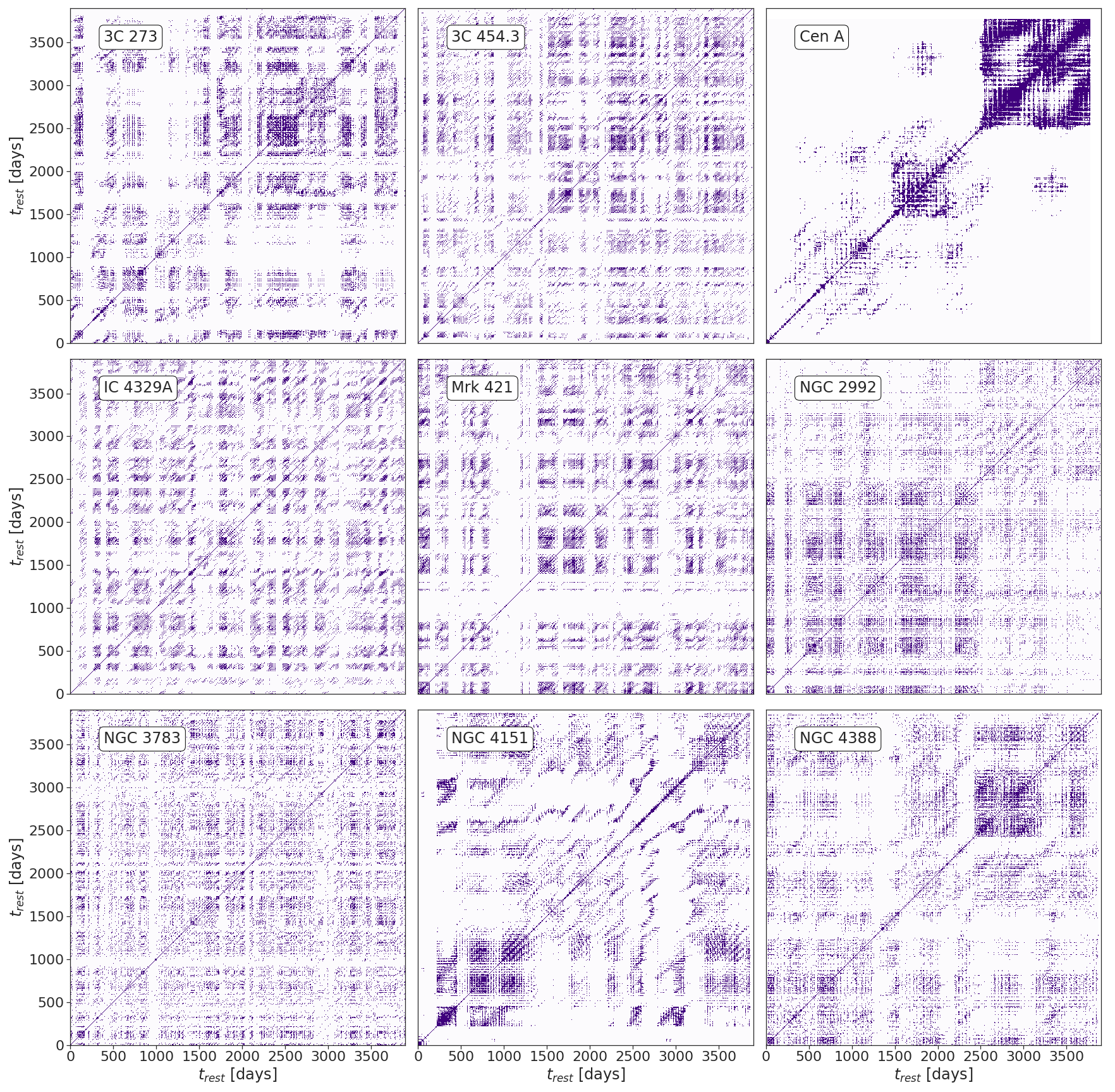}}
\caption{The recurrence plots of the 9 AGN plotted in Fig.~\ref{fig:AGN_LCs_select} for a threshold corresponding to a recurrence rate of 15 percent. Diagonally oriented features represent periodic or deterministic structure, vertical/horizontal lines indicate time-invariant states or systematics, large scale changes in structure indicate state changes or non-stationarity, and randomly distributed points not part of lines indicate randomness. For example, Cen A (top-right) demonstrates non-stationarity while NGC 4151 (bottom-center) exhibits changing texture indicative of possible state changing. Chopped up diagonal lines in all the RPs indicate stochastic or chaotic behavior.}
\label{fig:AGN_RPs_select} 
\end{figure*}

\subsubsection{Supplemental Tests for Non-stationarity}

RPs are useful for identifying nonlinearity in time series. However, the nonlinearity is degenerate with non-stationarity, in which the mean, variance, or autocorrelations of the time series are time-dependent. One of these time-dependent features occurs when a time series contains a ``unit root.'' In order to disentangle detections of nonlinearity from mere non-stationarity, we will consider two unit root stationarity tests: the Augmented Dickey Fuller (ADF) test and the Kwiatkowski-Phillips-Schmidt-Shin (KPSS) test. 

To explore how a unit root is identified in a time series, it is useful to consider the trend-cycle decomposition of a time series: $y_t = TD_t + z_t$, where $TD_t$ is some deterministic process and $z_t$ is a first-order autoregressive process such that $z_t = \psi z_{t-1} + \epsilon_t$, and $\epsilon_t$ is white noise. The characteristic polynomial of the autoregressive equation is $(1-\psi z)$, which has a root at $z=1/\psi$. For $|\psi| < 1$ we have a stationary process, for $|\psi| > 1$ we have a deviant nonstationary process, and for $\psi = 1$ we have a nonstationary, random walk with a so-called ``unit root.'' Unit root tests are based on testing the hypothesis that $\psi = 1$. The nonstationarity of the unit root can be remedied through differencing the time series (e.g., $y_{t+1} - y_t$), thereby eliminating some of the time-dependency of the statistical metrics. If a unit root persists in the differenced time series, then we say the time series is difference non-stationary (the time-dependency is more complicated or higher-order). Similarly, if the time-dependent trend term is also included and persists through differencing despite the lack of a unit root, then the time series contains a trend, called trend non-stationary. 

The ADF test can be used to determine the presence of a unit root in the (differenced) time series. The null hypothesis of the test states that the series has a unit root. A rejection of the null hypothesis means the time series is difference-stationary (the time dependence of the statistical metrics can be removed via differencing the time series). In a similar way, the KPSS test is used to determine the presence of a trend in the time series. The null hypothesis states that the series is trend stationary, and a rejection means that it is not stationary. We utilize the StatsModels \citep{Seabold2010} functionality for both stationarity tests implemented in Python to determine the difference and trend stationarity of all 46 light curves. Only one object was determined to be both difference and trend non-stationary (NGC 5252) and 17 sources were determined to be trend non-stationary (identified in Table~\ref{tab:table2}). Non-stationarity in the underlying dynamical process can also be seen in the RPs of some sources, as noted in Cen A and NGC 5252. All sources with evidence of non-stationarity of any kind are identified in Table~\ref{tab:table2} by a ``k'' (trend non-stationary), ``a'' (difference non-stationary), or ``c'' (non-stationary by the RP). It is important to take note of which sources contain non-stationarity when we determine the presence of nonlinearity with the RP techniques. Evidence of nonlinearity in stationary sources indicates the nonlinearity is in the underlying physical process. 

\subsection{Testing for Nonlinearity and Determinism}

\begin{figure}
\centering
\subfloat{
	\includegraphics[width=0.45\textwidth]{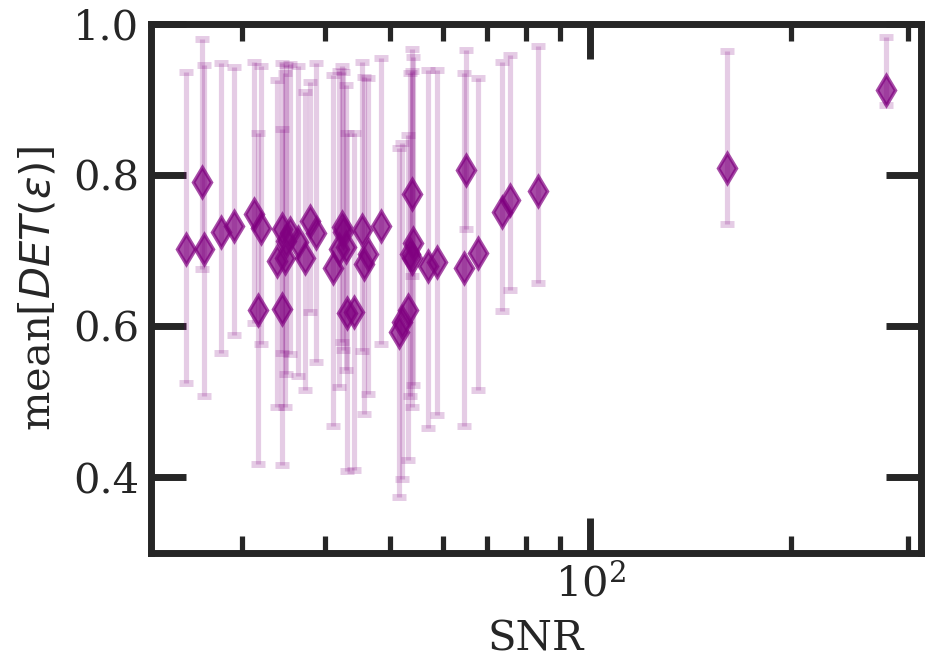}}\\
\subfloat{
	\includegraphics[width=0.45\textwidth]{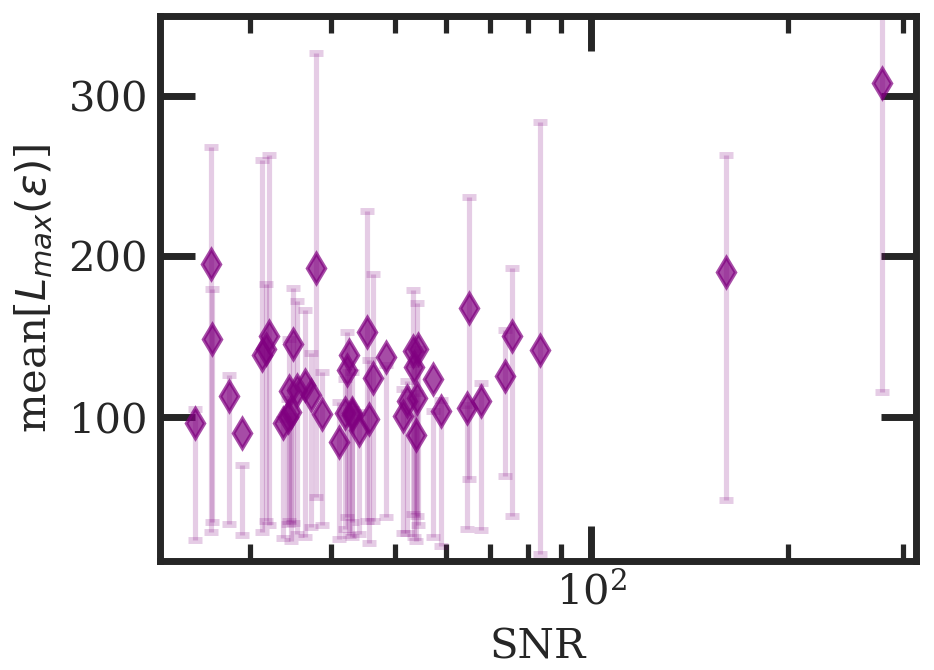}}\\
\subfloat{
	\includegraphics[width=0.45\textwidth]{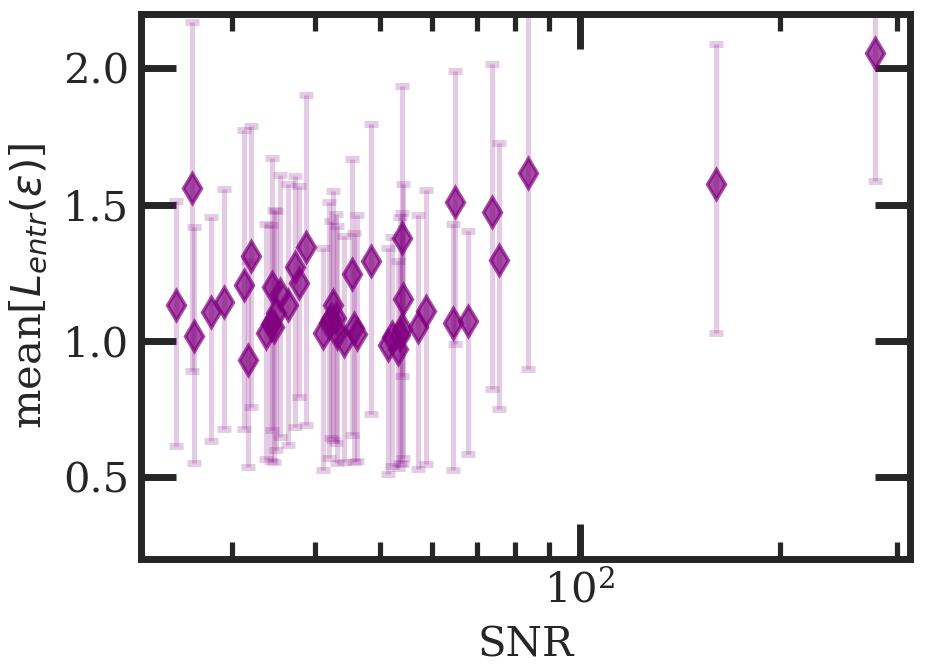}}
\caption{The average value of the determinism (DET; top), longest diagonal line length (Lmax; middle) and Shannon entropy (Lentr; bottom), for a recurrence rate between 5 percent and 40 percent, as a function of signal-to-noise for all AGN in our sample. The errorbars on each measurement represent the interquartile range of all possible RQA values that arise for thresholds corresponding to recurrence rates from 1 per cent to 99 per cent. There is no appreciable decline of the RQA measures for the lowest signal-to-noise sources. The two largest SNR sources correspond to NGC 4151 and Cen A.}
\label{fig:AGN_avgRQA_SNR} 
\end{figure}

While we can see, qualitatively, a diversity of RP behavior, we require a rigorous statistical test of the RP statistics compared to surrogate data in order to establish any correlations with source astronomical properties. We quantify the structure evident in the RPs of each source using three measures computed from the recurrence matrix (Eq.~\ref{eq:rp_mat1}): the longest diagonal line length ($L_{max}$) for the detection of nonlinearity, the determinism (DET) for the detection of determinism, and the Shannon entropy ($L_{entr}$) for the detection of information entropy, or uncertainty. We plot the average value and interquartile range of values for each recurrence measure versus signal-to-noise (as determined by the average flux in each 5-day light curve and average error) in Fig.~\ref{fig:AGN_avgRQA_SNR} in order to explore possible biases related to the brightness of an object. The spread of values for each source corresponds to the range of possible RQA values that occur depending on choice of threshold used in constructing the RP. We find that the 3 RQA measures used are not impacted strongly by the brightness of a source, regardless of choice of RP threshold, with a majority of the sample with a SNR between 20 and 100. 

For each object, we compute the 3 RQA measures for both the data RP and every set of surrogate data types. We compare these quantities against all surrogate types as a function of threshold. To determine the level of significance at a specific threshold, we use a rank-order significance test \citep{Theiler1992}. We select a residual probability $\alpha$ of a false rejection, corresponding to a significance $(1-\alpha) \times 100\%$ for a generated $M=K/\alpha -1$ surrogates. The probability that the data by coincidence has one of the largest values is exactly $\alpha$. For our given 100 surrogates, a 95 per cent confidence level that the RQA measure is significant versus a specific surrogate type would correspond to our data representing one of the 5 largest (or smallest) values of the $L_{max}$, $DET$, or $L_{entr}$ measures, for a given threshold. An example of determining the $L_{max}$ significance for Her X-1 against its surrogates is presented in Appendix A, available online. We will use the mean significance over a range of thresholds that correspond to a 5 per cent to 40 per cent recurrence rate to compare all of the AGN in our sample.

We also compute the time reversal asymmetry statistic and the simple nonlinear prediction error for all sources and their surrogates and use the rank-order test to determine significance against the surrogates. In Table~\ref{tab:table2}, we display the average significance against each type of surrogate for $L_{max}$, $DET$, and $L_{entr}$, as well as the time reversal asymmetry statistic (TR) and simple nonlinear prediction error (SNP).

We have several categories of general results for this sample:
\begin{itemize}
\item Overall, there is high significance of the $DET$ measure with respect to the phase surrogates. This result indicates that there is (likely non-periodic) recurrent structure in the \textit{Swift}/BAT light curves that is not characterized by the PSD alone.
\item Only a few objects contain high significance of nonlinearity as determined by the simple nonlinear prediction error (nonlinearity due to a deterministic process), while more objects (roughly half our sample) demonstrate time reversal asymmetry with respect to surrogates (nonlinearity due to a stochastic or deterministic process).
\item The small number of sources with significance in the $L_{max}$ measure with respect to the IAAFT surrogates indicates that there are only a few deterministic nonlinear light curves.
\item Mrk 6 is the only source that contains insignificance in all measures compared to all surrogate types, which indicates it is indiscernible from uncorrelated noise. This source is also in the bottom tier of signal to noise (from the HEASARC) in this sample.
\end{itemize}

\begin{table*}
\caption{The average significance of the determinism (DET), longest diagonal line length ($L_{max}$), and Shannon entropy ($L_{entr}$) for a range of thresholds corresponding to recurrence rates between 5 percent and 40 percent, the time reversal asymmetry statistic (TR), and simple nonlinear prediction error (SNP) against three types of surrogates (IAAFT - ``I'', phase - ``P'', and shuffled - ``S''). High significance of a measure is defined by $>$95 per cent confidence according to the rank-order test and is indicated by ``S'' marked with an asterisk. Those measures which are marginally significant ($>$90 per cent) are marked with simply ``S''. Objects with detected non-stationarity are annotated: ($k$) trend non-stationary, ($a$) difference non-stationary, and ($c$) dynamically non-stationary by the recurrence plot. The table is organized identically to Table~\ref{tab:table1} (first by spectroscopic type, then by radio properties).}
\begin{tabular}{l|lll|lll|lll|lll|lll} 
\hline
\multicolumn{16}{c}{Average Significance of RQA Measures against Surrogates}\\
\hline
\textbf{Object} & \multicolumn{3}{c}{DET} & \multicolumn{3}{c}{$L_{max}$} & \multicolumn{3}{c}{$L_{entr}$} & \multicolumn{3}{c}{TR} & \multicolumn{3}{c}{SNP} \\
& I & P & S & I & P & S & I & P & S & I & P & S & I & P & S \\
\hline
3C 273$^k$        &       & S$^*$ & S$^*$ &       & S     & S$^*$ &       & S$^*$ & S$^*$ &       & S$^*$ &       &  &  & \\
3C 454.3          & S$^*$ & S$^*$ & S$^*$ &       & S     & S$^*$ & S$^*$ & S$^*$ & S$^*$ &       & S     &       &  &  & \\
Mrk 421           & S$^*$ & S$^*$ & S$^*$ &       &       & S$^*$ & S     & S$^*$ & S$^*$ & S     & S$^*$ & S$^*$ &  & S & \\
\hline
3C 111            & S$^*$ & S$^*$ & S$^*$ & S     & S$^*$ & S$^*$ & S$^*$ & S$^*$ & S$^*$ & S$^*$ & S$^*$ & S$^*$ &  &  & \\
3C 382            &       & S     & S$^*$ &       &       &       &       & S     & S$^*$ &       &       &       &  &  & \\
4C 50.55          &       &       & S     &       &       &       &       &       & S$^*$ &       &       &       &  &  & \\
NGC 1275          & S$^*$ & S$^*$ & S$^*$ &       &       & S     & S$^*$ & S$^*$ & S$^*$ & S$^*$ & S$^*$ & S$^*$ &  &  & \\
QSO B0241+622     &       & S$^*$ & S$^*$ &       &       &       &       & S     & S$^*$ &       & S$^*$ & S$^*$ &  &  & \\
MCG+08-11-011$^k$ &       & S$^*$ & S$^*$ &       &       & S     &       & S$^*$ & S$^*$ &       &       &       &  &  & \\
NGC 3516$^k$      &       &       & S$^*$ & S     & S     & S$^*$ &       &       & S$^*$ &       &       &       &  &  & \\
NGC 5548$^k$      & S$^*$ & S$^*$ & S$^*$ &       &       &       & S$^*$ & S$^*$ & S$^*$ & S$^*$ & S$^*$ & S$^*$ & S$^*$ & S$^*$ & S$^*$\\
4U 0557-38        &       &       &       &       &       &       & S$^*$ & S$^*$ & S$^*$ &       &       &       &  & S & \\
GRS 1734-292      &       & S$^*$ & S$^*$ &       &       & S     & S     & S$^*$ & S$^*$ & S     & S$^*$ & S$^*$ &  & S$^*$ & \\
IC 4329A$^k$      &       & S     & S$^*$ &       &       &       & S$^*$ & S$^*$ & S$^*$ & S     & S$^*$ &       &  &  & \\
MR 2251-178       & S     & S$^*$ & S$^*$ &       &       &       & S     & S$^*$ & S$^*$ & S$^*$ & S$^*$ & S     &  &  & \\
Mrk 6             &       &       &       &       &       &       &       &       &       &       &       &       &  &  & \\
Mrk 110$^k$       & S     & S$^*$ & S$^*$ &       &       &       &       & S$^*$ & S$^*$ &       & S     &       &  &  & \\
Mrk 926           &       & S     & S$^*$ &       & S     & S$^*$ & S     & S$^*$ & S$^*$ &       &       &       &  &  & \\
NGC 3783          & S$^*$ & S$^*$ & S$^*$ &       &       &       & S$^*$ & S$^*$ & S$^*$ &       &       &       &  &  & \\
NGC 4151$^k$      & S     & S$^*$ & S$^*$ &       &       & S$^*$ & S$^*$ & S$^*$ & S$^*$ &       &       &       & S$^*$ & S$^*$ & S$^*$\\
NGC 4593          &       &       & S$^*$ &       &       & S$^*$ &       &       & S$^*$ &       &       &       &  &  & \\
NGC 6814$^k$      & S$^*$ & S$^*$ & S$^*$ & S     & S     & S     & S$^*$ & S$^*$ & S$^*$ & S$^*$ & S$^*$ & S$^*$ &  &  & \\
\hline
Cen A$^c$         &       &       & S$^*$ &       &       & S$^*$ &       &       & S$^*$ &       &       &       &  &  & \\
Cyg A$^k$         &       &       & S     &       &       &       & S$^*$ & S     & S$^*$ &       &       &       &  &  & \\
Mrk 348$^k$       & S$^*$ & S$^*$ & S$^*$ &       &       &       & S$^*$ & S$^*$ & S$^*$ &       & S$^*$ &       &  &  & \\
PKS 0018-19$^k$   & S$^*$ & S$^*$ & S$^*$ &       &       &       & S$^*$ & S$^*$ & S$^*$ &       &       &       &  &  & \\
NGC 2110          &       &       & S$^*$ &       &       & S$^*$ &       &       & S$^*$ & S$^*$ & S$^*$ & S$^*$ &  &  & \\
NGC 5728          &       &       & S$^*$ &       &       &       &       &       & S$^*$ &       & S$^*$ &       &  &  & \\
4U 1344-60        & S$^*$ & S$^*$ & S$^*$ &       &       &       & S$^*$ & S$^*$ & S$^*$ &       & S$^*$ &       &  &  & \\
Circinus Galaxy   & S     & S     & S$^*$ &       &       &       & S     & S     & S$^*$ &       & S     &       &  &  & \\
ESO 103-035       & S$^*$ & S$^*$ & S$^*$ &       &       &       & S$^*$ & S$^*$ & S$^*$ &       &       &       &  &  & \\
ESO 297-018       & S$^*$ & S$^*$ & S$^*$ & S     & S     &       & S$^*$ & S$^*$ & S$^*$ &       & S     &       &  &  & \\
ESO 506-027$^k$   & S$^*$ & S$^*$ & S$^*$ &       &       &       & S$^*$ & S$^*$ & S$^*$ &       &       &       & S & S & S$^*$\\
MCG-05-23-016     & S$^*$ & S$^*$ & S$^*$ &       &       &       & S$^*$ & S$^*$ & S$^*$ & S     & S$^*$ &       &  &  & \\
NGC 1365          & S$^*$ & S$^*$ & S$^*$ &       &       &       & S$^*$ & S$^*$ & S$^*$ &       &       &       &  &  & \\
NGC 2992$^k$      &       & S$^*$ & S$^*$ &       &       & S$^*$ & S$^*$ & S$^*$ & S$^*$ &       &       &       & S$^*$ & S$^*$ & S$^*$\\
NGC 3081$^k$      & S$^*$ & S$^*$ & S$^*$ &       &       &       & S$^*$ & S$^*$ & S$^*$ &       &       &       &  &  & \\
NGC 3281          & S$^*$ & S$^*$ & S$^*$ &       &       &       & S$^*$ & S$^*$ & S$^*$ & S$^*$ & S$^*$ & S$^*$ &  &  & \\
NGC 4388$^k$      &       &       & S$^*$ & S     & S     & S$^*$ & S     & S     & S$^*$ &       &       &       &  &  & \\
NGC 4507$^k$      &       &       &       &       &       &       & S$^*$ & S$^*$ & S$^*$ &       &       &       &  &  & \\
NGC 4945          &       &       &       &       &       & S     &       &       & S$^*$ &       &       &       &  &  & \\
NGC 5252$^{ack}$  & S$^*$ & S$^*$ & S$^*$ & S$^*$ & S$^*$ & S$^*$ & S$^*$ & S$^*$ & S$^*$ & S$^*$ & S$^*$ & S$^*$ &  &  & \\
NGC 5506          & S$^*$ & S$^*$ & S$^*$ &       &       &       & S$^*$ & S$^*$ & S$^*$ & S$^*$ & S$^*$ & S$^*$ &  &  & \\
NGC 7172          & S$^*$ & S$^*$ & S$^*$ &       &       &       & S$^*$ & S$^*$ & S$^*$ & S$^*$ & S$^*$ & S     &  &  & \\
NGC 7582          & S$^*$ & S$^*$ & S$^*$ &       &       &       & S$^*$ & S$^*$ & S$^*$ &       &       &       &  &  & \\
XSS J05054-2348$^k$ & S   & S$^*$ & S$^*$ &       &       &       &       & S$^*$ & S$^*$ &       & S$^*$ & S     &  &  &                                                  
\end{tabular}
\label{tab:table2}
\end{table*}

\section[Results]{Recurrence Correlations with Physical Properties}
\label{sec:results}

We compare the average significance of each of the RQA measures from Table~\ref{tab:table2} to the physical characteristics of each source from Table~\ref{tab:table1}. Specifically, we seek to determine whether there is an increased likelihood of a specific recurrence feature associated with different astronomical properties such as spectroscopic type, obscuration, radio loudness, or Giustini-Proga classification (based on a combination of the Eddington ratio and the black hole mass; \citealt{Giustini2019}). 

We consider two comparisons: the proportion of significance of each RQA measure (e.g., the fraction of Type 1 AGN that have a significant DET measure, labelled as `S' or `S$^*$' in Table~\ref{tab:table2}, compared to Type 2 AGN), and a comparison of the average of each RQA measure as listed in Table~\ref{tab:table2} (e.g., the mean determinism of Type 1 AGN versus Type 2 AGN). For comparing the proportions of significance, we use the Fisher exact test (alternative to the chi-square test for independence, but for comparing proportions of small sample sizes; \citealt{Fisher}). For comparing the mean RQA measures, we use the two-sided Welch's T-test (alternative to the standard t-test to test the hypothesis that two populations have equal means, but for small sampe sizes; \citealt{Welch1947}). 

A majority of the AGN in our sample have significant DET and $L_{entr}$ measures versus the phase and shuffled surrogates and thus we only consider the significance against the IAAFT surrogates for $DET$ and $L_{entr}$. Similarly, only a handful of sources showed significant nonlinearity according to the simple nonlinear prediction error test (ESO 506-027, NGC 2992, NGC 4151, and NGC 5548 against the IAAFT surrogates and 4U 0557-38, Mrk 421 against the phase surrogates) and so this measure is not compared between groups of astronomical properties. 

While the surrogate data method allows us to distinguish dynamical behavior from stochastic, the mean $DET$, mean $L_{max}$, and mean $L_{entr}$ provide a general picture of the characteristics of the recurrence plot. The mean recurrence features allow us to determine whether there are systematically higher features among any category of AGN. For example, a higher mean $DET$ tells us the recurrence plot contains more diagonal lines, a higher mean $L_{entr}$ indicates the distribution of diagonal lines is more complex, and a higher $L_{max}$ indicates a higher occurrence of persistent recurrent trends. A trend with any of these values does not indicate a distinction from stochastic surrogates --- an increase in any value could be consistent with a surrogate with the same PSD and flux distribution --- but it does indicate a trend with complex variability.

\subsubsection{Spectroscopic Type}

Overall, we find that 40 per cent of Type 1 AGN and 27 per cent of Type 2 AGN have significant time reversal asymmetry as compared to their IAAFT surrogates. These numbers are the same when we include the blazars in with the Type 1 AGN; however, the difference in proportions is not significant.

We find that 40 per cent of Type 1 AGN have significant determinism, compared to 70 per cent of Type 2 AGN, versus the IAAFT surrogates. We find the difference in proportions is modestly significant, with a p-value of 0.07 (via the Fisher test) when comparing Type 1 and Type 2. This suggests that the Type 2 AGN carry more frequent evidence of deterministic behavior compared to Type 1 AGN. Furthermore, the mean $DET$ measure over a range of thresholds for the Type 2 AGN is systematically higher than for Type 1, with a p-value of 0.007 (via the Welch's T-test) indicating this difference is highly significant.

\bigskip

\begin{center}
\begin{tabular}{l|lll}
\hline
\multicolumn{4}{c}{Fraction Significant w.r.t. IAAFT Surrogates}\\
\hline
     & Type 1 & Type 2 & p-value \\
\hline
$DET$      & 8/20   & 18/26  & 0.07    \\
$L_{max}$  & 3/20   & 3/26   & 0.9     \\
$L_{entr}$ & 11/20  & 21/26  & 0.1  \\
TR         & 8/20   & 7/19   & 0.53    
\end{tabular}
\captionof{table}{The fraction of Type 1 and Type 2 AGN that have significant determinism (DET), nonlinearity ($L_{max}$), or entropy ($L_{entr}$) relative to the IAAFT surrogates. The proportions in each population of AGN are compared using the Fisher exact test. The third column lists the p-value of rejecting the null hypothesis that the proportions are equal. In this case, Type 1 AGN contain a higher proportion of deterministic sources relative to Type 2 AGN, with a p-value of 0.07.}
\label{tab:type_fisher} 
\end{center}

\bigskip

When we consider the $L_{max}$ statistic, which measures traces of nonlinearity, we find no difference between Type 1 and Type 2 AGN, where 15 per cent of Type 1 and 12 per cent of Type 2 AGN carry significance versus the IAAFT surrogates. However, a comparison of the average $L_{max}$ measure between spectroscopic types does reveal a significant difference with a p-value of 0.014, suggesting a higher nonlinearity measure on average among Type 2 AGN. The inclusion of blazars with the Type 1 AGN effects the same results. The difference in the average entropy is also highly significant, with a p-value of 0.009. This suggests that Type 2 AGN contain systematically higher measures of information entropy (i.e., uncertainty in the distribution of the light curve) and topological entropy (i.e., higher-order variability measured by $L_{max}$). While these features are higher, they are not distinguishable from their surrogates and thus merely tell us that there is greater complexity to the light curves in Type 2 AGN. The significance is not impacted by including blazars with the Type 1 AGN. 

In summary, Type 2 AGN are more likely to carry determinism with respect to the IAAFT surrogates (which tells us the traces of determinism in the RP are not stochastic in origin) compared to Type 1 AGN, with systematically higher values of the recurrence features generally. This tells us that Type 2 AGN are more likely to contain higher-order features of variability and evidence of non-stochastic behavior than Type 1 AGN. The dynamical differences between these populations suggest a relationship between the accretion process and the detection of the emission lines that dictate spectroscopic class, which could challenge the traditional unification model of AGN that type is a factor of viewing angle if this relationship persists in a larger sample. 

\bigskip

\begin{center}
\begin{tabular}{l|lll}
\hline
\multicolumn{4}{c}{Mean RQA Measures}\\
\hline
          & Type 1 & Type 2 & p-value \\
\hline
mean $DET$  & 0.743    & 0.763    & 0.007    \\
mean $L_{max}$ & 28.97    & 32.61    & 0.014     \\
mean $L_{entr}$ & 1.336    & 1.431    & 0.009    
\end{tabular}
\captionof{table}{The average value of the determinism (DET), nonlinearity ($L_{max}$), or entropy ($L_{entr}$) for Type 1 and Type 2 AGN. The means of the recurrence measures in each population of AGN are compared using the two-sided Welch's T-test. The third column lists the p-value of rejecting the null hypothesis that the means are equal. In this case, Type 1 AGN contain lower values for all recurrence features relative to Type 2 AGN.}
\end{center}

\bigskip

\subsubsection{Obscuration effects}

We find that there are no significant differences in the proportions of AGN with significance against the IAAFT surrogates for any of the RQA measures or the time reversal asymmetry statistic among the obscured and unobscured AGN. Similarly, none of the RQA measures are detected as systematically higher for one population over the other. The most significant result, with a p-value of 0.1, is for a slightly higher determinism measure among obscured AGN, but this is not considered highly significant.

We therefore conclude that both populations have equal likelihood to be distinct from stochastic noise, and any differing recurrence features are not due to a nonlinear or deterministic process. Overall, we find minimal correlations between the dynamics of the hard X-ray variability and obscuration properties. 

\bigskip

\begin{center}
\begin{tabular}{l|lll}
\hline
\multicolumn{4}{c}{Fraction Significant w.r.t. IAAFT Surrogates}\\
\hline
     & Obs. & Unobs. & p-value \\
\hline
$DET$ & 16/25   & 10/21  & 0.37    \\
$L_{max}$ & 3/25   & 3/21   & 0.9     \\
$L_{entr}$ & 19/25  & 13/21  & 0.35    \\
TR         & 7/25   & 8/21   & 0.54  
\end{tabular}
\captionof{table}{The fraction of obscured (Obs.) and unobscured (Unobs.) AGN that have significant determinism (DET), nonlinearity ($L_{max}$), or entropy ($L_{entr}$) relative to the IAAFT surrogates. The proportions in each population of AGN are compared using the Fisher exact test. The third column lists the p-value of rejecting the null hypothesis that the proportions are equal. In this case, there are no significant differences in the proportions of significant recurrence features that are distinct from the surrogates between obscured and unobscured AGN.}
\label{tab:obs_fisher} 
\end{center}

\bigskip

\begin{center}
\begin{tabular}{l|lll}
\hline
\multicolumn{4}{c}{Mean RQA Measures}\\
\hline
     & Obs. & Unobs. & p-value \\
\hline
mean $DET$  & 0.759   & 0.748  & 0.10    \\
mean $L_{max}$ & 31.91   & 29.98   & 0.19     \\
mean $L_{entr}$ & 1.413  & 1.361  & 0.14    
\end{tabular}
\captionof{table}{The average value of the determinism (DET), nonlinearity ($L_{max}$), or entropy ($L_{entr}$) for obscured (Obs.) and unobscured (Unobs.) AGN. The means of the recurrence measures in each population of AGN are compared using the two-sided Welch's T-test. In this case, the average determinism among unobscured AGN is slightly lower than obscured AGN, with a p-value of 0.1.}
\end{center}

\subsubsection{Radio properties}

Radio observations have found compact radio emission in the majority of radio-quiet AGN (e.g., \citealt{Nagar2002}; \citealt{Panessa2010}; \citealt{Maini2016}) and parsec-scale extended jet-like morphologies have previously been observed in some radio quiet AGN (e.g., \citealt{Orienti2010}). This has led to the theory that radio quiet AGN are in fact scaled down versions of their radio loud counterparts with unresolved jet-like morphologies. \cite{Smith2020} found general consistency among a sample of 70 \textit{Swift}/BAT-selected AGN observed with the Karl G. Jansky Very Large Array 22 GHz radio imaging with predictions of scale-invariant jet models for the origin of radio emission in radio-quiet AGN (and also with predictions of coronal models). Five of the AGN in our sample were classified by \cite{Smith2020} to have miniature jet-like radio emission and are identified in Table~\ref{tab:table1}. When comparing the recurrence properties between radio quiet and radio loud AGN, we include the jet-like radio quiet AGN with the traditional radio loud objects, in order to test for the role of possible jet influence on hard X-ray variability. Indeed, correlations have been observed between the radio and X-ray luminosities in AGN of all types, which may be related to the supermassive black hole mass in a ``fundamental plane of black hole activity'' \citep{Merloni2003}. 

We find no significant difference between radio loud and radio quiet sources when we measure their time reversal asymmetry statistic relative to any of the surrogate types.
There are also no differences in the average RQA measures between radio loud and radio quiet AGN. However, 83 per cent of radio quiet AGN have significant entropy relative to their IAAFT surrogates compared to 47 per cent of radio loud AGN. This difference is highly significant, with a p-value of 0.02. We note that though the average RQA measures are not significantly different to a p-value of 0.05, the radio quiet sources do contain marginally higher values. We therefore conclude that radio quiet AGN are much more likely to contain uncertainty in the generating distribution of the light curve, and this effect is distinguishable from stochastic light curves, and they exhibit generally more complex variability.

In summary, we find that radio quiet AGN are more likely to carry higher information entropy compared to surrogates with the same PSD and flux distribution. This indicates higher-order information (either high-dimensional stochastic or chaotic) is contained in the light curves of radio quiet AGN that is not recoverable from the PSD alone. We can infer that the accretion processes that dominate in radio quiet sources lead to more chaotic behavior while a strong radio source (i.e., presence of a jet) suppresses variability and uncertainty.

\bigskip

\begin{center}
\begin{tabular}{l|lll}
\hline
\multicolumn{4}{c}{Fraction Significant w.r.t. IAAFT Surrogates}\\
\hline
     & RQ & RL & p-value \\
\hline
$DET$  & 19/29   & 7/17  & 0.13    \\
$L_{max}$ & 4/29   & 2/17   & 0.9     \\
$L_{entr}$ & 24/29  & 8/17  & 0.02    \\
TR         & 9/29  & 6/17  &  0.9  
\end{tabular}
\captionof{table}{The fraction of radio quiet (RQ) and radio loud (RL) AGN that have significant determinism (DET), nonlinearity ($L_{max}$), or entropy ($L_{entr}$) relative to the IAAFT surrogates. The proportions in each population of AGN are compared using the Fisher exact test. The third column lists the p-value of rejecting the null hypothesis that the proportions are equal. In this case, radio quiet AGN contain a higher proportion of AGN with significant entropy relative to radio loud sources, with a p-value of 0.02.}
\label{tab:radio_fisher} 
\end{center}

\bigskip

\begin{center}
\begin{tabular}{l|lll}
\hline
\multicolumn{4}{c}{Mean RQA Measures}\\
\hline
          & RQ       & RL       & p-value \\
\hline
mean $DET$  & 0.759    & 0.748    & 0.12    \\
mean $L_{max}$ & 32.07    & 29.67    & 0.09     \\
mean $L_{entr}$ & 1.414    & 1.358    & 0.1     
\end{tabular}
\captionof{table}{The average value of the determinism (DET), nonlinearity ($L_{max}$), or entropy ($L_{entr}$) for radio quiet (RQ) and radio loud (RL) AGN. The means of the recurrence measures in each population of AGN are compared using the two-sided Welch's T-test. The third column lists the p-value of rejecting the null hypothesis that the means are equal. In this case, the means of the recurrence properties are the same between radio quiet and radio loud AGN.}
\end{center}

\subsubsection{Giustini-Proga Classification}\label{sec:GPclass}

We consider the classifcations of AGN type established by \cite{Giustini2019} based on combinations of low/high accretion rate and low/high black hole mass. These classifications are akin to the high-intensity/soft-energy and low-intensity/hard-energy spectral states of Galactic black hole X-ray binaries. In summary, there are 6 possible states outlined by \cite{Giustini2019}:
\begin{itemize}
\item State 1: very low accretion rate ($L/L_{Edd}$ less than $10^{-6}$)
\item State 2: low accretion rate ($10^{-6} < L/L_{Edd} < 10^{-3}$)
\item State 3: ``moderate'' accretion rate (up to 0.25 $L/L_{Edd}$), and less than $10^8$ $M_{sol}$
\item State 4: moderate accretion rate, and greater than $10^8$ $M_{sol}$
\item State 5: ``high'' accretion rate (up to unity $L/L_{Edd}$), and less than $10^8$ $M_{sol}$
\item State 6: high accretion rate, and greater than $10^8$ $M_{sol}$
\item State 7: very high accretion rate ($L/L_{Edd}$ much greater than 1)
\end{itemize}
In our sample of \textit{Swift}/BAT AGN we find no sources in state 1, 9 in state 2, 17 in state 3, 11 in state 4, 7 in state 5, and one each in states 6 and 7. We therefore define four different classes: sources in states 2, 3, 4, and 5-7 combined, which gives us a low accretion rate class, a high accretion rate class, and two moderate accretion rate classes with a distinction based on black hole mass. We compare every pair of classes and determine whether there is a significant difference between the RQA measures relative to the surrogates, or in the average values of each RQA measure.

Overall, there appear to be no differences in the significance among the different Giustini-Proga classes for the $DET$ or $L_{entr}$ measures. In measuring significant nonlinearity, we find that over a third of the sources in the `state 4' classification have a significant detection of nonlinearity relative to their surrogates compared to none in the `state 2' classification. The difference in these proportions is modestly significant to a p-value of 0.09. The primary distinction between these two classes is the accretion rate, as the `state 2' class has no criterion based on black hole mass but contains all the low accretion rate sources, while the `state 4' class has high mass objects with moderate accretion rates. 

\bigskip

\begin{center}
\begin{tabular}{l|lll}
\hline
\multicolumn{4}{c}{Fraction Significant w.r.t. IAAFT Surrogates}\\
\hline
     & State 2 & State 4 & p-value \\
\hline
$DET$   & 6/9   & 9/11  & 0.62    \\
$L_{max}$ & 0/9   & 4/11   & 0.09     \\
$L_{entr}$ & 6/9  & 10/11  & 0.28   
\end{tabular}
\captionof{table}{The fraction of Giustini-Proga `state 2' and `state 4' AGN that have significant determinism (DET), nonlinearity ($L_{max}$), or entropy ($L_{entr}$) relative to the IAAFT surrogates. The proportions in each population of AGN are compared using the Fisher exact test. The third column lists the p-value of rejecting the null hypothesis that the proportions are equal. In this case, the state 2 (low Eddington ratio) AGN contain a higher proportion of AGN with significant nonlinearity relative to the state 4 (low mass, moderate Eddington ratio) AGN, with a p-value 0.09.}
\label{tab:gp34_fisher} 
\end{center}

\bigskip

\begin{center}
\begin{tabular}{l|lll}
\hline
\multicolumn{4}{c}{Mean RQA Measures}\\
\hline
     & State 2 & State 4 & p-value \\
\hline
mean $DET$  & 0.762   & 0.744  & 0.07    \\
mean $L_{max}$ & 32.20   & 28.96   & 0.15     \\
mean $L_{entr}$ & 1.423  & 1.338  & 0.08    
\end{tabular}
\captionof{table}{The average value of the determinism (DET), nonlinearity ($L_{max}$), or entropy ($L_{entr}$) for Giustini-Proga `state 2' and `state 4' AGN. The means of the recurrence measures in each population of AGN are compared using the two-sided Welch's T-test. In this case, the average determinism and average entropy are both higher among state 2 AGN (low Eddington ratio) compared to the state 4 AGN (low mass, moderate Eddington ratio), with a p-value of 0.07 and 0.08, respectively.}
\end{center}

\bigskip

The distinction between the `state 2' and `state 4' classifications persists when we consider the average RQA values. The low accretion rate objects have a systematically higher nonlinearity measure (p-value 0.09), higher determinism (p-value 0.07), and higher entropy (p-value 0.07). Since these comparisons are with p-values greater than 0.05, we do not consider these effects to be highly significant. The same trend persists when comparing the `state 3' to `state 4' sources, but is not significant (p-values 0.14, 0.18, and 0.16 for higher nonlinearity, determinism, and entropy, respectively, in the state 3 objects). This may indicate that the combination of higher mass and higher accretion rate may translate to fewer recurrence features, while those features carry higher-order information than what is contained in the PSD, but a much larger sample is needed to explore these relationships further.

In summary, we find that high mass, moderate accretion rate systems have a higher proportion of significant nonlinearity as compared to the lowest accretion rate objects, albeit with modest significance, while the latter contain more variable RPs (though this variability is not distinguishable from stochastic variability).

\subsubsection{Black Hole Mass, Bolometric Luminosity, and Eddington Ratio}

We compare the recurrence properties (their significance relative to surrogates and their mean values) to black hole mass, bolometric luminosity, and Eddington ratio using the Spearman correlation coefficient \citep{Spearman1904}. Overall, we find no significant correlations between recurrence properties and any of the physical properties for the whole sample. The strongest relationship we find is an anti-correlation between bolometric luminosity and each of the mean RQA measures, though the correlation is weak (-0.27 each with insigificant p-values). Given that there appear to be different variability characteristics between classes of AGN (e.g., Type 1 versus Type 2), we determine if any correlations exist within each category. 

As a preliminary step, we perform an Anderson-Darling test on each pair of samples (e.g., Type 1 versus Type 2) to test the null hypothesis that the distributions of each category of AGN are consistent in black hole mass, Eddington ratio, bolometric luminosity, and SNR. This is to determine whether there are selection biases that potentially impact any observed differences between categories of AGN. When comparing spectroscopic types and obscuration properties, we find the null hypothesis that the samples in each category are drawn from the same distribution cannot be rejected in terms of mass, Eddington ratio, luminosity, or SNR. That is, differences in the variability characteristics in these samples are not due to differences in any of these properties. However, our samples of radio loud and radio quiet AGN are distinct in their distributions of black hole mass and bolometric luminosity to a p-value of 0.05 in both cases, where the radio loud sources typically have higher ranges of values. That is the radio quiet AGN have log $L_{bol} = [42.73 - 45.44]$ and log $M_{bh} = [6.15 - 8.99]$ while the radio loud AGN have log $L_{bol} = [43.26 - 48.58]$ and log $M_{bh} = [6.48 - 9.4]$. Differences in the populations of radio quiet versus radio loud AGN variability characteristics may be related to these samples containing offset values in mass and luminosity; although, given the large errors at least in the former, this may not be a strong caveat and could merely reinforce the premise that variability indeed correlates with astronomical properties.

We compute the Spearman correlation coefficient to determine the correlations between black hole mass, bolometric luminosity, and Eddington ratio within each category of AGN (Type 1, Type 2, obscured, unobscured, radio quiet, and radio loud), as we did for the entire sample. For the Eddington ratio, we consider low accretion rate systems ($\lambda < 0.2$) separate from high accretion rate systems. The results of these correlations are summarized in Tables~\ref{tab:corr_edd} through~\ref{tab:corr_lum}. 

Recurrence analysis is most useful for distinguishing features relative to surrogates, allowing for distinctions between stochastic and deterministic behaviors. We find no significant relationships with black hole mass and any category of AGN that distinguish variability from stochastic surrogates. We do find a relationship between radio properties and Eddington ratio and bolometric luminosity. Radio loud AGN contain a positive correlation between deterministic behavior and Eddington ratio, with a coefficient of 0.55 and a p-value of 0.05. This is our most significant result among all features distinct from the IAAFT surrogates. The radio loud AGN also contain a correlation between significant levels of information entropy and both Eddington ratio and bolometric luminosity, with coefficients of 0.54 and 0.45 and p-values of 0.06 and 0.07, respectively. Finally, there is a positive correlation between significant nonlinearity and bolometric luminosity. 

To summarize: the accretion rate and bolometric luminosity of an AGN greatly impact the detection of variability that is distinct from surrogates when a radio jet is present in the system. An increase in accretion rate results in more distinct determinism, while an increase in luminosity results in more distinct nonlinearity; and an increase in both result in greater entropy. This tells us that the jet and accretion process are coupled, and that the jet additionally suppresses stochastic variability.

The greater number of correlations exist when we consider the general features of the recurrence plot. When considering Eddington radio, there are differences in the mean recurrence features and each set of AGN characteristics. Type 1, unobscured, and radio loud AGN all contain anti-correlations with Eddington ratio and all of these trends are highly significant except for the mean determinism of Type 1 AGN (p-value of 0.1). An example of the mean entropy as a function of Eddington ratio is displayed in Fig.~\ref{fig:AGN_RQA_vs_props}, top panel. Bolometric luminosity and black hole mass have impacts on the opposite categories: obscured and radio quiet AGN exhibit anti-correlations with mean recurrence features, though all of these relationships are not highly significant, with p-values ranging between 0.06 and 0.1. If we consider a combination of Eddington ratio and black hole mass by looking at the different Giustini-Proga (GP) classes, we find there may be a relationship between GP classes and the mean recurrence features. For example, Fig.~\ref{fig:AGN_RQA_vs_props} bottom panel displays the distribution of the mean entropy across three GP classes, where the highest entropy values tend to congregate in the fourth GP class of low-mass, moderate accretion rate systems. These correlations could be more effectively studied among a larger sample of AGN.

In summary, the radio properties of AGN alone dictate whether we observe an increase in distinct and non-stochastic variability as a function of accretion rate or luminosity. This infers a connection of the radio jet to X-ray variability where for high values of the accretion rate and luminosity, there is more deterministic modulation of the variability. Secondly, Type 1, unobscured, and radio loud AGN all exhibit less complex variability as a function of accretion rate while Type 2, obscured, and radio quiet sources are unaffected by accretion rate. This infers a connection between the classification of Seyfert type and associated obscuration to the accretion properties of these systems: Type 1 and unobscured AGN variability are sensitive to the accretion rate. All of these correlations reinforce the differences between Type 1 and Type 2 AGN, obscured and unobscured AGN, and between radio-loud and radio-quiet AGN and additionally suggest that the accretion rate in particular affects each set of populations differently.

It should be noted that though the Spearman correlation can be computed for sample sizes as low as 10 (e.g., \citealt{Zar1972}), the reliability of the p-values is by no means robust for our small sample. By eye, these relationships are weak and thus a larger sample (e.g., the entire catalog of $Swift$/BAT AGN) would be required to assert a relationship with any physical property.

\begin{center}
\includegraphics[width=0.45\textwidth]{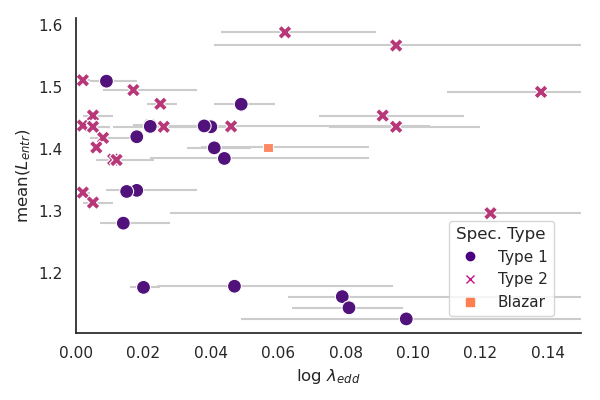}\\
\includegraphics[width=0.45\textwidth]{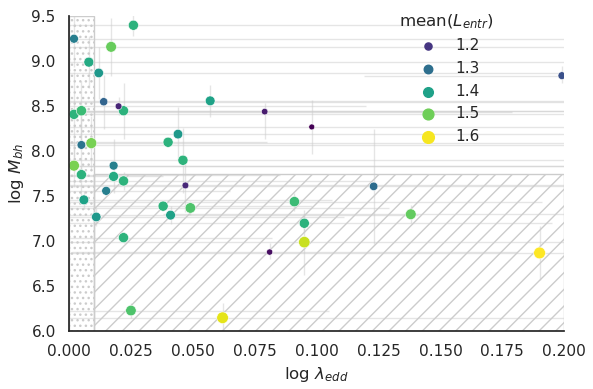}
\captionof{figure}{Top: the average $L_{entr}$ measure for Type 1 AGN (purple solid circles), Type 2 AGN (light purple crosses), and blazars (orange solid squares) as a function of Eddington ratio, where the errors are plotted in solid grey lines. Bottom: the black hole mass as a function of Eddington ratio in which the color and size of the markers indicate the mean $L_{entr}$ measure (error bars in grey). The dotted hash indicates the Giustini-Proga state 2 (low accretion rate), the lined hash indicates the Giustini-Proga state 4 (low mass, moderate accretion rate), and the un-hashed section corresponds to Giustine-Proga state 3 (high mass, moderate accretion rate). The $L_{entr}$ of the Type 1 AGN carry an anti-correlation with Eddington ratio while the Type 2 AGN do not, and the higher accretion rate, lower-mass AGN show a higher $L_{entr}$ on average compared to the other Giustini-Proga states.}
\label{fig:AGN_RQA_vs_props} 
\end{center}

\begin{table*}
\begin{tabular}{l|lll|lll}
\hline
\multicolumn{7}{c}{Correlations with Eddington Ratio}\\
\hline
 & DET Significance & Lmax Significance & ENTR Significance & Mean DET      & Mean Lmax     & Mean ENTR     \\
 \hline
Type 1 & -0.02 (0.93)     & -0.09 (0.72)      & -0.21 (0.44)      & \textbf{-0.43 (0.10)}  & \textbf{-0.57 (0.02)}  & \textbf{-0.59 (0.02)}  \\
Type 2 & 0.19 (0.41)      & 0.19 (0.40)       & 0.27 (0.22)       & 0.25 (0.26)   & 0.22 (0.31)   & 0.20 (0.37)   \\
Obs.   & 0.21 (0.36)      & 0.20 (0.37)       & 0.28 (0.20)       & 0.21 (0.34)   & 0.27 (0.23)   & 0.24 (0.28)   \\
Unobs. & -0.05 (0.86)     & -0.15 (0.57)      & -0.20 (0.44)      & \textbf{-0.72 (0.001)} & \textbf{-0.69 (0.002)} & \textbf{-0.70 (0.002)} \\
RQ     & -0.23 (0.27)     & -0.15 (0.47)      & -0.28 (0.17)      & 0.06 (0.79)   & 0.18 (0.39)   & 0.12 (0.57)   \\
RL     & \textbf{0.55 (0.05)}      & 0.41 (0.16)       & \textbf{0.54 (0.06)}       & \textbf{-0.59 (0.03)}  & \textbf{-0.59 (0.04)}  & \textbf{-0.59 (0.03)} 
\end{tabular}
\captionof{table}{The Spearman correlation coefficient between the significance of each recurrence feature relative to the IAAFT surrogates --- determinism (DET), nonlinearity ($L_{max}$), and entropy ($L_{entr}$) --- and Eddington ratio in columns 1-3, and the correlation between the average recurrence measure and Eddington ratio in columns 4-6. The correlation coefficient is listed first, with the p-value of the correlation in parentheses. For example, the radio loud (RL) AGN have a positive correlation significant $DET$ with accretion rate, with a coefficient of 0.55 and p-value of 0.05 (the determinism of radio loud AGN become more significant relative to IAAFT surrogates with increasing Eddington ratio). All correlations with p-values of 0.1 or less are highlighted in boldface.}\label{tab:corr_edd}
\end{table*}

\begin{table*}
\begin{tabular}{l|lll|lll}
\hline
\multicolumn{7}{c}{Correlations with Black Hole Mass}\\
\hline
 & DET Significance & Lmax Significance & ENTR Significance & Mean DET      & Mean Lmax     & Mean ENTR     \\
 \hline
Type 1  & -0.01 (0.96) & 0.03 (0.91)  & 0.16 (0.50)  & -0.29 (0.21) & -0.33 (0.16)  & -0.31 (0.18)  \\
Type 2  & -0.09 (0.66) & -0.07 (0.75) & -0.17 (0.43) & -0.20 (0.36) & -0.25 (0.25)  & -0.23 (0.28) \\
Obs.    & -0.14 (0.50) & -0.06 (0.76) & -0.19 (0.36) & -0.30 (0.14) & \textbf{-0.36 (0.08)}  & \textbf{-0.34 (0.10)} \\
Unobs.  & 0.19 (0.42)  & 0.09 (0.71)  & 0.24 (0.29)  & 0.02 (0.92)  & -0.001 (0.99) & 0.01 (0.95) \\
RQ      & 0.22 (0.24)  & 0.16 (0.41)  & 0.26 (0.18)  & -0.29 (0.12) & \textbf{-0.36 (0.06)}  & \textbf{-0.33 (0.08)} \\
RL      & -0.06 (0.83) & -0.03 (0.90) & 0.02 (0.95)  & 0.26 (0.31)  & 0.27 (0.30)   & 0.27 (0.29)                  
\end{tabular}
\captionof{table}{The Spearman correlation coefficient between the significance of each recurrence feature relative to the IAAFT surrogates --- determinism (DET), nonlinearity ($L_{max}$), and entropy ($L_{entr}$) --- and Black Hole Mass in columns 1-3, and the correlation between the average recurrence measure and Black Hole Mass in columns 4-6. The correlation coefficient is listed first, with the p-value of the correlation in parentheses. For example, the radio quiet (RQ) AGN have a negative correlation of mean $L_{max}$ with mass, with a coefficient of -0.36 and p-value of 0.06 (the average $L_{max}$ value decreases with increasing mass). All correlations with p-values of 0.1 or less are highlighted in boldface.}
\label{tab:corr_mass}
\end{table*}

\begin{table*}
\begin{tabular}{l|lll|lll}
\hline
\multicolumn{7}{c}{Correlations with Bolometric Luminosity}\\
\hline
    & DET Significance & Lmax Significance & ENTR Significance & Mean DET      & Mean Lmax    & Mean ENTR     \\
\hline
Type 1 & -0.04 (0.88)     & 0.003 (0.99)      & 0.12 (0.61)       & -0.37 (0.11)  & -0.36 (0.12) & \textbf{-0.37 (0.10)}  \\
Type 2 & 0.15 (0.47)      & 0.11 (0.59)       & 0.14 (0.52)       & -0.25 (0.25)  & -0.27 (0.21) & -0.26 (0.22)  \\
Obs.   & 0.08 (0.97)      & 0.04 (0.84)       & 0.07 (0.73)       & \textbf{-0.34 (0.10)}  & \textbf{-0.38 (0.06)} & \textbf{-0.37 (0.07)}  \\
Unobs. & 0.25 (0.27)      & 0.19 (0.38)       & 0.24 (0.29)       & -0.013 (0.95) & 0.001 (0.99) & -0.004 (0.98) \\
RQ     & 0.07 (0.72)      & -0.09 (0.64)      & 0.09 (0.64)       & \textbf{-0.35 (0.06) } & \textbf{-0.35 (0.06)} & \textbf{-0.36 (0.06) } \\
RL     & 0.35 (0.17)      & \textbf{0.44 (0.08)}       & \textbf{0.45 (0.07)}       & 0.08 (0.77)   & 0.09 (0.74)  & 0.08 (0.76)  
\end{tabular}
\captionof{table}{The Spearman correlation coefficient between the significance of each recurrence feature relative to the IAAFT surrogates --- determinism (DET), nonlinearity ($L_{max}$), and entropy ($L_{entr}$) --- and Bolometric Luminosity in columns 1-3, and the correlation between the average recurrence measure and Bolometric Luminosity in columns 4-6. The correlation coefficient is listed first, with the p-value of the correlation in parentheses. For example, the radio loud (RL) AGN have a positive correlation of significant entropy with bolometric luminosity with a coefficient of 0.45 and p-value of 0.07 (the nonlinearity of radio loud AGN become more significant relative to IAAFT surrogates with increasing luminosity). All correlations with p-values of 0.1 or less are highlighted in boldface.}\label{tab:corr_lum}
\end{table*}

\section{Discussion and Conclusions} \label{sec:conclusions}

For a sample of 46 AGN monitored in the hard X-ray (14-150 keV) by \textit{Swift}/BAT, containing Seyfert 1s, Seyfert 2s, and 3 blazars, we have generated recurrence plots and quantitative recurrence measures that illustrate a broad range of variability characteristics in the phase space structure of the light curves. In the RQA analysis, most objects are highly significant against phase surrogates, which retain the same PSD as the source light curves. At a minimum we can conclude there is significant structure in the light curves that is not unveiled through second order metrics. 

We find that approximately half of the sources contain significant time reversal asymmetry in their light curves distinct from surrogates with the same PSD. Time irreversibility can be produced from both deterministic nonlinear systems, such as non-equilibrium systems, as well as nonlinear stochastic systems, such as nonlinear shot noise models (e.g., where the decay of the shot is nonlinear rather than linear; \citealt{Eliazar2005}). \cite{Maccarone2002} propose a shot noise model containing an asymmetric envelope of shots that sufficiently explains the time skewness evident in two X-ray binaries. Time irreversibility evident in many AGN in our sample rules out all gaussian processes and static transformations of a linear Gaussian process as appropriate models for their observed variability and support models that involve a nonlinear impulse-response system.

We find that four sources contain evidence for deterministic non-linearity according to a standard surrogate data test (simple nonlinear prediction error): ESO 506-027, NGC 2992, NGC 4151, and NGC 5548, all of which were identified as non-stationary from the ADF and KPSS non-stationarity tests. From the recurrence plot analysis we additionally found evidence for non-linearity in 3C 111, ESO 297-018, NGC 3516, NGC 4388, NGC 5252, and NGC 6814. Of these nonlinear sources, two were not identified to contain non-stationarity (3C 111 and ESO 297-018) and thus the detection of this non-linearity may in fact be due to a nonlinear, non-stochastic underlying process generating the emission. 

Over half of the light curves (10 non-stationary and 16 considered stationary) contain significant determinism relative to their IAAFT surrogates (retaining the same PSD and flux distribution). This result implies the origin of the determinism is truly non-stochastic. Similarly, a majority of the AGN (30 total) contain significant information entropy in their light curves relative to their IAAFT surrogates, indicating high information content and uncertainty in the distribution beyond what is contained in the PSD. In addition to high-order information, high information entropy combined with levels of determinism is also a probe of chaos \citep{Zanin2021}. The prevalence of high entropy appears irrespective of non-stationary behavior. To reiterate: the detections of significant determinism and entropy are distinguishable from stochastic variability for a large fraction of the sources we study.

When comparing the significance of three RQA measures relating to nonlinearity, determinism, and entropy to astronomical properties, we find that Type 2 AGN are significantly more likely to be deterministic compared to Type 1 AGN, and radio quiet AGN are more likely to contain higher information entropy compared to radio loud AGN. Both of these results are found significant relative to surrogates with the same PSD and flux distribution. It should be noted that previous studies of the \textit{Swift}/BAT AGN (e.g., \citealt{Beckmann2007} and \citealt{Soldi2014}) found that absorbed Type 2 AGN showed more variability than unabsorbed Type 1 AGN, and radio loud sources are the most variable. We confirm the findings that distinguish Type 1 and Type 2 AGN and further add that the source of the variability among Type 2 AGN is possibly related to a deterministic process, and the variability among radio loud AGN may be chaotic in origin or contain highly dimensional stochastic uncertainty. We did not find differences between obscured and unobscured AGN based on average features of their RPs or their comparison to stochastic surrogates. We therefore conclude any variability differences between obscured and unobscured AGN is not due to an underlying deterministic or nonlinear process and may be decoupled from the classification of Seyfert type by emission line features.

Finally, we compare the significance of each RQA measure with individual physical characteristics of each source: the bolometric luminosity, Eddington ratio (defined as $L_{bol}/L_{Edd}$), and black hole mass. We found no significant correlations between the RQA measures and their physical characteristics when considering the full sample of AGN; the strongest relationship is an anti-correlation between the average RQA measures and bolometric luminosity (i.e., decreasing complex variability with luminosity). This implies that low-luminosity AGN contain more complex (less noise-like) recurrence plots. When we consider individual classes of AGN, we find several weak correlations. Only radio loud AGN contained a relationship between the significance of non-stochastic behavior and Eddington ratio or bolometric luminosity.

There are several significant correlations when we consider the correlation between general recurrence plot features (defined as the mean of each of the RQA measures) and each of the physical properties. The most significant (with p-values less than 0.05) are related to the Eddington ratio. We find anti-correlations of the recurrence features with accretion rate for Type 1, unobscured, and radio loud AGN, while no correlations exist for the counterparts to these categories. We can infer that higher accretion rate sources have less complex recurrence plots, particularly among Type 1, unobscured, and radio loud AGN. It is notable that there were no significant differences between obscured and unobscured AGN when considering only their recurrence properties; but the introduction of accretion rate reveals a dependency of these features on obscuration properties.

In summary, we find that the recurrence plot contains significant information about the light curves of AGN beyond what is contained in the PSD and related techniques. This suggests that the power-laws evident in the PSDs of AGN do not provide information on the dynamical diversity in AGN light curves. In particular, it is clear that the recurrence plots of Type 1 AGN are significantly different from those of Type 2 AGN, an effect which is impacted by the accretion rate in these systems. It is notable that the difference between obscured and unobscured AGN is not as strong as that between Seyfert type. In fact, the RPs of obscured AGN contain features that correlate with black hole mass and luminosity while those of unobscured AGN correlate with accretion rate. 

This result is on its face initially confusing, given that the hard X-rays of the inner accretion flow should penetrate obscuration effects. The implication is that the hard X-ray variability is related to, for example, the equivalent widths of emission lines used for classifying Seyfert type and not on the obscuration properties, which suggests Seyfert type is not merely contingent on viewing angle. Furthermore, the variability characteristics are not merely due to a difference in the variance of the light curves. If Type 2 AGN variability is typically more deterministic, this suggests the inner accretion flow is dynamically distinct from that of Type 1 AGN.

Previous studies have similarly challenged the traditional unification theory of AGN. For example, \cite{Ricci2011} finds that though Type 1 and Type 2 AGN were found to have the same average nuclear continuum emission, the reflection component was significantly stronger for Compton-thin Seyfert 2s than for Seyfert 1s. One explanation for this observation is that the differences in the average reflection spectra of Seyfert 1s and Seyfert 2s arises from significant differences in their torus covering factors. This interpretation is supported by the statistical framework that Type 1 and Type 2 AGN are preferentially drawn from separate ends of the distribution of torus covering factors \citep{Elitzur2012}. 
Among the \textit{Swift}/BAT AGN sample, \cite{Hinkle2021} recovers a relationship between the reflection coefficient and photon index that supports the idea that soft spectra sources exhibit stronger reflection (e.g., as posited by \citealt{Zdziarski1999}) and finds that Type 2 AGN tend to have lower Eddington ratios than Type 1 AGN. And finally, \cite{Rojas2020} finds a higher occurrence of ionized outflows in Type 1.9 and Type 1 AGN with respect to Type 2 AGN and a significant dependence between outflow occurrence and accretion rate, which becomes relevant at Eddington ratios above 0.02. These observations support the idea that the current and historical outflowing properties of an AGN are related to the covering factor of the torus and, furthermore, are related to the reflection component of the AGN that appears in the hard X-rays. 

We observe differing recurrence properties of the Type 1 and Type 2 AGN in our sample but we do not detect as strong a difference in recurrence properties between obscured and unobscured AGN. Furthermore, accretion rate impacts the recurrence features of Type 1 and Type 2 differently, where an anti-correlation between recurrence features and Eddington ratio is seen among Type 1 AGN and not among Type 2 AGN. The fact that there are differences in the recurrence properties irrespective of obscuration but impacted by accretion rate indicates intrinsic hard X-ray variability due to the dynamics of the accretion flow and coronal emission itself. This supports the idea that Type 2 and Type 1 AGN sample from different populations of accretion flow properties: stronger reflection, lower accretion rates, and low occurrence of outflows among Type 2 and the opposite among Type 1, in which the latter leads to variability indistinct from stochastic noise. A similar distinction between recurrence features and accretion states is observed among microquasars \citep{Sukova2016}. The observation that there are distinct dynamical classes of the accretion process between these more complex definitions of AGN types could support the results from studies of changing-look AGN (e.g., \citealt{Denney2014}, \citealt{Liu2022}). Alternatively, \cite{Arur2020} have found that nonlinear properties of the variability in XRBs depend on inclination angle, and present a model that explains the dependency through changes on the optical depth of the coronal region over precessional timescales. Such a model would connect the hard X-ray emitting inner accretion flow to the extended coronal atmosphere of the disk or torus and explain variations in accretion flow properties such as reflection, accretion rates, and outflows for different spectroscopic types of AGN.

As a final note, the radio loud sources exhibit distinct variability from their radio quiet counterparts and should likely be treated as a separate sample. It is the clear the role of the radio jet critically changes the hard X-ray variability characteristics. In order to further explore the relationships between recurrence properties and astronomical properties, we will perform a follow-up study involving all of the AGN detected above a 5 signal-to-noise ratio in the 157-month catalog of \textit{Swift}/BAT in combination with the second data release of the physical properties of 857 AGN in the sample obtained by the BASS collaboration.

\section*{Acknowledgements}
R.A.P. is grateful for support from the NASA Education Minority University Research Education Project (MUREP) through the NASA Harriett G. Jenkins Graduate Fellowship activity under Grant No. 80NSSC19K1291. R.A.P. also acknowledges support for this work provided by the National Science Foundation MPS-Ascend Postdoctoral Research Fellowship under Grant No. 2138155. The authors thank the anonymous referee for comments that substantially improved the manuscript.

\section*{Data Availability}

This research made use of data retrieved from the publicly available repository of the Swift Burst Alert Telescope 157-month catalog (https://swift.gsfc.nasa.gov/results/bs157mon). The code used to analyze the data are the publicly available packages TISEAN (\citealt{Hegger1999}, \citealt{Schreiber2000}; available at https://www.pks.mpg.de/$\sim$tisean) for the production of surrogates and computation of the time reversal asymmetry statistic and simple nonlinear prediction error, and PyUnicorn (\citealt{Donges2015}; available at http://www.pik-potsdam.de/$\sim$donges/pyunicorn) for the production of RPs and RQA measures. The combination of these packages for use in the analyses in this paper was facilitated by scripts written in Python and will be shared on reasonable request to the corresponding author.

\bibliography{References.bib}

\end{document}